\documentclass[onecolumn]{aa}
\usepackage{graphicx}
\usepackage{txfonts}

\begin{document}

\title{{\it XMM--Newton} observations of a sample of $\gamma-$ray loud active galactic nuclei\thanks{Based 
on public observations obtained with \emph{XMM--Newton}, an ESA science mission with instruments and 
contributions directly funded by ESA Member States and the USA (NASA).}}

\author{L. Foschini\inst{1}, G. Ghisellini\inst{2}, C.M. Raiteri\inst{3}, F. Tavecchio\inst{2}, M. Villata\inst{3}, 
L. Maraschi\inst{2}, E. Pian\inst{4}, G. Tagliaferri\inst{2}, G. Di Cocco\inst{1}, G. Malaguti\inst{1}}

\institute{INAF/IASF-Bologna, Via Gobetti 101, 40129, Bologna (Italy)
\and
INAF, Osservatorio Astronomico di Brera, Via Bianchi 46, 23807, Merate (Italy)
\and
INAF, Osservatorio Astronomico di Torino, Via Osservatorio 20, 10025, Pino Torinese (Italy)
\and
INAF, Osservatorio Astronomico di Trieste, Via G.B. Tiepolo 11, 34131, Trieste (Italy)}

\offprints{\texttt{foschini@iasfbo.inaf.it} (L. Foschini).}
\date{Received 27 January 2006; Accepted 9 March 2006}

\abstract
{}
{To understand the nature of $\gamma-$ray loud active galactic nuclei (AGN) and the mechanisms for the 
generation of high-energy $\gamma-$rays.}
{We performed a homogeneous and systematic analysis of simultaneous X-ray and optical/UV properties of a group of 
$15$ $\gamma-$ray loud AGN, using observations performed with \emph{XMM-Newton}. The sample is composed of 13 blazars 
(6 BL Lac and 7 Flat-Spectrum Radio Quasar) and 2 radio galaxies that are associated with detections at energies 
$>100$~MeV. The data for 7 of them are analyzed here for the first time, including the first X-ray observation of 
PKS~$1406-706$. The spectral characteristics of the sources in the present sample were compared with 
those in previous catalogs of blazars and other AGN, to search for difference or long term changes.}
{All the selected sources appear to follow the classic ``blazar sequence'' and the spectral energy distributions 
(SED) built with the present X-ray and optical/UV data and completed with historical data, confirm the findings of previous 
studies on this type of source. Some sources display interesting features: four of them, namely 
AO~$0235+164$, PKS~$1127-145$, S5~$0836+710$ and PKS~$1830-211$ show the presence of an intervening absorption 
system along the line of sight, but only the last is known to be gravitationally lensed. AO~$0235+164$ 
was detected during an outburst and its SED shows a clear shift of the synchrotron peak. 
3C~$273$ shows a change in state with respect to the previous \emph{BeppoSAX} observations that can be 
interpreted as an increase of the Seyfert-like component and a corresponding decline of the jet emission. This is 
consistent with the monitoring at radio wavelengths performed during the same period. PKS~$1406-706$ is detected 
with a flux higher than in the past, but with a corresponding low optical flux. Although it is classified 
as FSRQ, the SED can be modelled with a simple synchrotron self-Compton model.}
{}

\authorrunning{L. Foschini et al.}

\maketitle

\section{Introduction}
There is general consensus on the supermassive black hole (SMBH) paradigm as the central engine of active galactic
nuclei (AGN). It is more difficult to obtain quantitative understanding of the physical mechanisms responsible 
for the observed properties of these cosmic sources. In the AGN zoo, $\gamma-$ray loud objects -- where with
this term we consider the AGN detected at $E>100$~MeV --  represent a small,
but interesting class. Their discovery dates back to the start of $\gamma-$ray astronomy, when 
the European satellite \emph{COS-B} ($1975-1982$) detected photons in the $50-500$~MeV range from 3C273 (Swanenburg et al. 1978). 
However, 3C273 remained the only AGN detected by \emph{COS-B}. A breakthrough in this research field came later with
the Energetic Gamma Ray Experiment Telescope (EGRET) on board the \emph{Compton Gamma-Ray Observatory} (\emph{CGRO}, 1991-2000).
The third catalog of point sources contains $271$ sources detected at energies greater than $100$~MeV: 
$93$ of them are identified with blazars ($66$ at high confidence and $27$ at low confidence), and $1$ with the nearby 
radiogalaxy Centaurus~A (Hartman et al. 1999). Among the remaining sources, there are $5$ pulsars, the Large Magellanic 
Cloud, one exceptional solar flare, and $170$ are unidentified. Therefore, EGRET discovered that the blazar-type AGN 
are the primary source of the extragalactic background in the MeV-GeV range, as suggested 
by several authors (e.g. Strong et al. 2004, Giommi et al. 2006).

\begin{table*}[!ht]
\caption{
Main characteristics of the observed AGN. Columns: (1) Name of the source from the Third EGRET Catalog; (2) Name of the
known counterpart; (3) Other name; (4) classification of the active nucleus (LBL: low frequency peaked BL Lacertae Object;
HBL: high frequency peaked BL Lacertae Object; FSRQ: flat-spectrum radio quasar; RG: radio galaxy); (5) Coordinates (J2000); 
(6) Redshift; (7) Galactic absorption column density [$10^{20}$~cm$^{-2}$] from Dickey \& Lockman (1990).}
\centering
\begin{tabular}{llllccc}
\hline
3EG          & Counterpart     & Other Name  & AGN Type & $\alpha$,$\delta$        & $z$    & $N_{\mathrm{H}}$ \\
(1)          & (2)             & (3)         & (4)      & (5)                      & (6)    & (7)              \\
\hline
J$0222+4253$ & $0219+428$      & 3C~$66$A    & LBL      & $02:22:39.6$,$+43:02:08$ & $0.444$  &  $8.99$         \\
J$0237+1635$ & AO~$0235+164$   &             & LBL      & $02:38:38.9$,$+16:36:59$ & $0.94$   &  $8.95$         \\
J$0530-3626$ & PKS~$0521-365$  &             & FSRQ     & $05:22:58.0$,$-36:27:31$ & $0.05534$&  $3.33$         \\
J$0721+7120$ & S5~$0716+714$   &             & LBL      & $07:21:53.4$,$+71:20:36$ & $>0.5^{\mathrm{*}}$   &  $3.81$         \\
J$0845+7049$ & S5~$0836+710$   & 4C~$71.07$  & FSRQ     & $08:41:24.3$,$+70:53:42$ & $2.172$  &  $2.91$         \\
J$1104+3809$ & Mkn~$421$       &             & HBL      & $11:04:27.3$,$+38:12:32$ & $0.03002$&  $1.38$         \\
J$1134-1530$ & PKS~$1127-145$  &             & FSRQ     & $11:30:07.0$,$-14:49:27$ & $1.184$  &  $4.02$         \\
J$1222+2841$ & ON~$231$        & W~Comae     & LBL      & $12:21:31.7$,$+28:13:59$ & $0.102$  &  $1.88$         \\
J$1229+0210$ & 3C~$273$        &             & FSRQ     & $12:29:06.7$,$+02:03:09$ & $0.15834$&  $1.79$         \\
J$1324-4314$ & Cen A           & NGC~$5128$  & RG       & $13:25:27.6$,$-43:01:09$ & $0.00182^{\mathrm{**}}$ &  $8.62$	        \\
J$1339-1419$ & PKS~$1334-127$  &             & FSRQ     & $13:37:39.8$,$-12:57:25$ & $0.539$  &  $4.82$         \\
J$1409-0745$ & PKS~$1406-076$  &             & FSRQ     & $14:08:56.5$,$-07:52:27$ & $1.494$  &  $2.77$         \\
J$1621+8203$ & NGC~$6251$      &             & RG       & $16:32:32.0$,$+82:32:16$ & $0.0247$ &  $5.47$         \\
J$1832-2110$ & PKS~$1830-211$  &             & FSRQ     & $18:33:39.9$,$-21:03:40$ & $2.507$  &  $21.9$         \\
J$2158-3023$ & PKS~$2155-304$  &             & HBL      & $21:58:52.0$,$-30:13:32$ & $0.116$  &  $1.69$         \\
\hline
\end{tabular}
\begin{list}{}{}
\item[$^{\mathrm{*}}$] Lower limit evaluated on the basis of the non-detection of the host galaxy (Sbarufatti et al. 2005).
\item[$^{\mathrm{**}}$] This redshift is not indicative and the distance of $3.84$~Mpc is adopted here. See Evans et al.
(2004) for more details.
\end{list}
\label{tab:host}
\end{table*}

During the years following these discoveries, much effort has been dedicated to the identification of the 
remaining $170$ sources. This is a challenging enterprise given the large position error contours of EGRET sources
(typically $\sim 0^{\circ}.5-1^{\circ}$). 
A significant advancement has been obtained by Sowards-Emmerd et al. (2003, 2004), who performed a radio survey at
$8.4$~GHz. By using a ``figure of merit'' obtained combining the $8.4$~GHz flux, the radio spectral index
and the X-ray flux (when available), they proposed $20$ new identifications of EGRET sources with blazar-type AGN. 
A strong improvement in the identification and discovery of new $\gamma-$ray loud AGN is expected with the forthcoming
missions \emph{GLAST}\footnote{\texttt{http://www-glast.stanford.edu}} and 
\emph{AGILE}\footnote{\texttt{http://agile.iasf-milano.inaf.it}}.

An important complement to these discoveries was the observations performed by the Italian-Dutch satellite
\emph{BeppoSAX} (1996-2002) and operating in the $0.1-300$~keV energy band 
(see Ghisellini 2004 for a review on blazar observations with this satellite). During its lifetime, \emph{BeppoSAX} 
observed more than $80$ blazars (Giommi et al. 2002, Donato et al. 2005) and sampled with high sensitivity the X-ray 
region of the spectral energy distribution (SED) over more than three decades in energy. 

The observations of these and other high-energy satellites, together with ground telescopes, led to the discovery
that the spectral energy distribution (SED) of blazars is typically composed of two peaks, one due to synchrotron emission and the other to inverse 
Compton radiation, the latter discovered by \emph{CGRO}/EGRET (von Montigny et al. 1995). 
Maraschi et al. (1995b) and Sambruna et al. (1996) noted that the broad-band spectra of BL Lac and Flat-Spectrum 
Radio Quasars (FSRQ) share common features and properties (that justified the common designation of ``blazars''
proposed by Spiegel in 1978). Fossati et al. (1998) and Ghisellini et al. (1998) proposed a unified scheme where
the blazars are inserted into a ``sequence'' according to their physical characteristics. 
Low luminosity BL Lac have the synchrotron peak in the UV-soft X-ray energy band and therefore are ``high-energy
peaked'' (HBL). As the synchrotron peak shifts to low energies (near infrared, ``low-energy peaked'' BL Lac or LBL), 
the luminosity increases and the X-ray emission can be due to synchrotron or inverse Compton or a mixture of both.
In the case of FSRQ, the blazars with the highest luminosity, the synchrotron peak is in the far infrared and the 
X-ray emission is due to inverse Compton radiation. However, other authors reported failures in the above~mentioned scheme
(see, for example, Padovani et al. 2003, Landt et al. 2006). 

The two-peak SED is a dynamic picture of the blazar behaviour: indeed, these AGN are characterized by strong 
flares during which the SED can change dramatically (e.g. Tagliaferri et al. 2002). The most striking example 
of such a behaviour is represented by Mkn~$501$ -- although at $\gamma-$ray energies this source was not 
detected by EGRET, but by TeV telescopes -- that during an outburst in 1997 showed a shift of the synchrotron peak to 
the hard X-ray energy band ($50-100$~keV, Pian et al. 1998). 
The variability on different time scales and, particularly, the intraday variability, is one of the striking 
characteristics of blazars and is considered one of the proofs that the continuum is generated by a relativistic jet 
with a small observing angle (for a review see Wagner \& Witzel 1995, Ulrich et al. 1997). 

\emph{XMM-Newton} (launched in Dec. 1999, Jansen et al.~2001) is a satellite with the current largest collecting area, 
useful for timing studies, together with a good sensitivity and spectral 
resolution. \emph{XMM-Newton} covers a lower frequency range than \emph{BeppoSAX}, i.e. from the optical/UV domain to 
the X-rays, up to $10$~keV. From the large \emph{XMM-Newton} public archive, we have selected and analyzed 
all the publicly available EGRET-detected AGN data. Although the single observations were originally intended for other
purposes, it was possible to carry out a homogeneous analysis and to study the main spectral characteristics of these
sources in the X-ray energy band. Several sources of the present sample deserve particular attention and 
many detailed studies have been published on the individual sources. However, the aim of the present work is 
to perform an overall view and comparison with previous surveys to search for common features
that could explain the $\gamma-$ray generation. Some early resuls of the present work have been presented 
in Foschini et al. (2006).

This paper is organized as follows: in Sect.~2 the selection criteria, biases and the parameters used in the data 
analysis are presented; in Sect.~3, the spectral characteristics are shown and compared with other catalogs. 
The absorption systems are discussed in Sect.~4, and a short note on the X-ray spectral features is presented 
in Sect.~5. The spectral energy distributions and the blazar sequence is shown in Sect.~6. Sect.~7 contains final 
remarks. In Appendix A we provide the notes on the individual sources together with the fits and some tabular material.

The cosmology values adopted through the paper, when not explicitly declared, are 
$H_0=70$~km$\cdot$s$^{-1}$Mpc$^{-1}$, $\Omega_{\lambda}=0.7$ and $\Omega_{\rm m}=0.3$.

\section{Sample selection and data analysis}
The starting sample consists of all the AGN in the Third EGRET Catalog (Hartman et al. 1999) updated with the recent
results by Sowards-Emmerd et al. (2003, 2004). This sample has been cross-correlated with the public observations available 
in the \emph{XMM-Newton} Science Archive\footnote{\texttt{http://xmm.vilspa.esa.es/external/xmm\_data\_acc/xsa/index.shtml}} 
to search for spatial coincidences in the field of view (FOV) of the EPIC camera, within $10'$ of the 
boresight\footnote{This maximum distance has been selected to take into account that within that region the 
telescope vignetting is well corrected according to Kirsch~(2005).}. $15$ AGN have been found (Table~\ref{tab:host}) as 
of January $4^{\rm th}$, 2006, and for three of them there are more than 5 observations available 
(see the observation log in Table~\ref{tab:log}), making it possible also to study the long term behaviour.

\subsection{Biases and caveats}
The present work suffers from several biases, but nevertheless it is possible to obtain useful
information about the overall behaviour of the $\gamma-$ray loud AGN.

The first source of bias is the Third EGRET catalog itself: the large point spread function (PSF) of the 
EGRET telescope and its moderate sensitivity, changes in the position from the 2EG to the 3EG catalog, double 
(or more?) sources not resolved by the EGRET PSF (see the notes on the single sources). Sowards-Emmerd et al. (2005) 
called for the release of a Fourth 
EGRET catalog, but a major advancement will be possible when the \emph{GLAST} satellite is operational.  
The arcminute-sized PSF of the LAT telescope and the higher sensitivity would then improve the confidence of 
the suggested associations (e.g. 3EG~J$0530-3626$ or 3EG~J$1621+8203$) or disentangle the multiple contributions 
of certain EGRET sources (see, for example, 3EG~J$0222+4253$). 

The use of the \emph{XMM-Newton} public data introduces new biases. Exposures and instrument modes were not selected 
for a survey, but with completely different purposes (e.g. calibration, ToO, ...). In the 
\emph{XMM-Newton} archive there are many more pointings than the $46$ reported here: some were discarded because of 
problems in the processing of the observation data files (ODF), some others were not used because the instrument mode 
does not match the major part of the pointing of the same source\footnote{This is the case of Mkn 421, 3C 273, and 
PKS 2155-304, that are calibration sources. For the purposes of the present work, only the pointings with the PN 
detector in small window mode were used.}, or because the observations were still covered by the PI proprietary 
data rights at the time when the archive was scanned. 
In spite of $46$ observations, the present sample is made of only $15$ AGN. Three sources 
dominate: 3C~$273$ with $15$ observations, PKS~$2155-304$ with $9$, and Mkn~$421$ with $6$ (see 
the observation log in Table~\ref{tab:log}).  

The large differences in settings of the instrument modes between the individual observations 
had a particularly severe impact on the OM data (Table~\ref{tab:omdata}). It is not possible to have one 
single filter to be used as a reference for all the observations. In the best case, the magnitudes with filter UVW1 are 
available for $18$ of $46$ pointings. In one case only (PKS~$1830-211$), the optical counterpart of the blazar 
has a V magnitude $\approx 25$ (Courbin et al. 2002) and therefore is beyond the instrument capabilities.

\begin{table*}[!t]
\caption{Best fit model parameters for the 15 AGN (present work). In the case of sources with multiple pointings, 
the weighted averages are shown. Since PKS~$2155-304$ (3EG~J$2158-3023$) is best fitted in 4 pointings with the 
broken power law model and in the remaining 5 with the simple power law model; we reported the averages of both models.
Columns: (1) Name of the source; (2) absorbing column density [$10^{20}$~cm$^{-2}$]; (3) photon index
$\Gamma$, if the best fit is a simple power law model or soft photon index $\Gamma_1$, if the best fit
model is a broken power law; (4) hard photon index $\Gamma_2$ for the broken power law model; (5) Break energy [keV].}
\centering
\begin{tabular}{lcccc}
\hline
Name         & $N_{\rm H}$     & $\Gamma$/$\Gamma_1$    & $\Gamma_2$             & $E_{\rm break}$ \\
(1)          & (2)             & (3)                    & (4)                    & (5)             \\
\hline
$0219+428$    & Gal.            & $2.91_{-0.08}^{+0.12}$ & $2.23_{-0.09}^{+0.10}$ & $1.3\pm 0.2$\\
AO~$0235+164$ & Gal.            & $2.33\pm 0.04$         & $2.1\pm 0.1$           & $3.3_{-0.5}^{+0.7}$\\
PKS~$0521-365$& Gal.            & $1.95\pm 0.03$         & $1.74\pm 0.03$         & $1.5_{-0.2}^{+0.3}$\\
S5~$0716+714$ & Gal.            & $2.70\pm 0.02$         & $1.98_{-0.09}^{+0.08}$ & $2.3_{-0.1}^{+0.2}$\\
S5~$0836+710$ & $14\pm 3$       & $1.379\pm 0.007$       & $-$                    & $-$\\
Mkn~$421$     & Gal.            & $2.38\pm 0.09$         & $2.7\pm 0.2$           & $2.7\pm 1.0$\\
PKS~$1127-145$& $12_{-1}^{+2}$  & $1.40_{-0.05}^{+0.08}$ & $1.22\pm 0.06$         & $2.7_{-0.8}^{+1.0}$\\
ON~$231$      & $2.5\pm 0.6$    & $2.77\pm 0.04$         & $-$                    & $-$\\
3C~$273$      & Gal.            & $2.02\pm 0.08$         & $1.67\pm 0.05$         & $1.44\pm 0.08$\\
Cen~A         & $1523\pm 261$   & $2.22\pm 0.06$         & $-$                    & $-$\\
PKS~$1334-127$& $6.7\pm 0.9$    & $1.80\pm 0.04$         & $-$                    & $-$\\
PKS~$1406-076$& Gal.            & $1.59\pm 0.01$         & $-$                    & $-$ \\
NGC~$6251$    & $14\pm 1$       & $2.11_{-0.06}^{+0.08}$ & $1.78\pm 0.07$         & $2.5_{-0.4}^{+0.3}$\\
PKS~$1830-211$& $63\pm 1$       & $1.00\pm 0.09$         & $1.32\pm 0.06$         & $3.5\pm 0.7$\\
PKS~$2155-304$ & $1.69\pm 0.06$  & $2.9\pm 0.1$           & $-$                    & $-$ \\
{}             & Gal.            & $2.7\pm 0.1$           & $2.94\pm 0.06$         & $2.7\pm 0.7$ \\
\hline
\end{tabular}
\label{tab:averages}
\end{table*}

\begin{table*}[!h]
\caption{Best fit model parameters for the 15 AGN from the \emph{BeppoSAX} catalogs by Giommi et al.~(2002) and 
Donato et al.~(2005). The values are the weighted averages obtained from multiple observations (if any) and
from the two catalogs. In the case of 3C~$273$ (3EG~J$1229-0210$) there were 9 observations, 5 best fitted with a broken
power law and 4 with a single power law model: both model averages are reported. For NGC~$6251$ (3EG~J$1621+8203$) the 
average is obtained from Chiaberge et al. (2003) and Guainazzi et al. (2003) from a fit in the energy band $0.1-200$~keV. 
Grandi et al. (2003) is the reference for Cen~A (3EG~J$1324-4314$), from a fit in the energy band $0.4-250$~keV excluding
the $5.5-7.5$~keV band (iron emission line). The value for PKS~$1830-211$ (3EG~J$1832-2110$) has been obtained by De Rosa
et al. (2005) from a fit in the energy band $0.5-80$~keV (\emph{Chandra} and \emph{INTEGRAL}).
Columns: (1) Name of the source; (2) absorbing column density [$10^{20}$~cm$^{-2}$]; (3) photon index
$\Gamma$, if the best fit is a simple power law model or soft photon index $\Gamma_1$, if the best fit
model is a broken power law; (4) hard photon index $\Gamma_2$ for the broken power law model; (5) Break energy [keV].}
\centering
\begin{tabular}{lcccc}
\hline
Name          & $N_{\rm H}$     & $\Gamma$/$\Gamma_1$    & $\Gamma_2$             & $E_{\rm break}$ \\
(1)          & (2)             & (3)                    & (4)                    & (5)             \\
\hline
$0219+428$   & Gal.            & $2.22\pm 0.06$         & $-$                    & $-$\\
AO~$0235+164$ & Gal.            & $2.0\pm 0.1$           & $-$                    & $-$\\
PKS~$0521-365$ & Gal.            & $1.74\pm 0.02$         & $-$                    & $-$\\
S5~$0716+714$ & Gal.            & $2.5\pm 0.2$           & $1.8\pm 0.1$           & $3.0\pm 0.4$\\
S5~$0836+710$ & $78_{-35}^{+55}$& $1.31\pm 0.02$         & $-$                    & $-$\\
Mkn~$421$     & Gal.            & $1.9\pm 0.2$           & $2.3\pm 0.3$           & $1.3\pm 0.8$\\
PKS~$1127-145$ & Gal.            & $1.42\pm 0.05$         & $-$                    & $-$\\
ON~$231$      & Gal.            & $2.58\pm 0.01$         & $1.52\pm 0.06$         & $2.8\pm 0.2$\\
3C~$273$      & Gal.            & $2.0\pm 0.1$           & $1.603\pm 0.006$       & $0.9\pm 0.3$\\
{}            & Gal.            & $1.58\pm 0.03$         & $-$                    & $-$\\
Cen~A         & $1020_{-40}^{+90}$  & $1.80_{-0.04}^{+0.03}$ & $-$                    & $-$\\
PKS~$1334-127$ & $-$             & $-$                    & $-$                    & $-$\\
PKS~$1406-076$ & $-$             & $-$                    & $-$                    & $-$ \\
NGC~$6251$     & $9\pm 1$        & $1.79\pm 0.06$         & $-$                    & $-$\\
PKS~$1830-211$ & $194_{-25}^{+28}$ & $1.09\pm 0.05$       & $-$                    & $-$\\
PKS~$2155-304$ & Gal.            & $2.3\pm 0.1$           & $2.8\pm 0.1$           & $1.7\pm 0.2$ \\
\hline
\end{tabular}
\label{tab:bepposax}
\end{table*}

\subsection{Common procedures of data analysis}
For the processing, screening, and analysis of the data from the EPIC MOS1, MOS2 (Turner et al. 2001) and PN cameras
(Str\"uder et al. 2001), standard tools have been used (\texttt{XMM SAS v. 6.1.0} and \texttt{HEAsoft v 6.0}). The 
standard procedures described in Snowden et al. (2004) were followed. Only single pixel events have
been selected, excluding border pixels or columns with higher offset. High-background flares affected
the observations randomly, and in some cases it was necessary to filter the available data. Time intervals contaminated by
flares have been excluded by extracting the whole detector lightcurve for $E>10$~keV and by removing the periods with
count rates higher than $1.0$~s$^{-1}$ for PN and $0.35$~s$^{-1}$ for MOS, as suggested by Kirsh (2005).

The source spectra have been extracted from a circular region with a radius of $40''$ and centered in the catalog 
(radio) position of the AGN. The background to be subtracted in the analysis was derived from a circular region, with the 
same radius, near the selected source. In the case of pile-up, the source region is an annulus with still an external 
radius of $40''$ and an internal radius selected to minimize the pile-up effects by using the task \texttt{epatplot} of 
\texttt{XMM SAS}. It resulted $8''$ for 3C~$273$ and PKS~$2155-304$, and $10''$ for Mkn~$421$. For Cen~A, the annulus 
has internal and external radii of size $20''$ and $50''$, respectively. 

Since PN is the most stable detector, with negligible degradation of performance to date, we adopted it as the prime 
instrument. Data from MOS cameras have been used only when it was necessary to check a finding 
obtained with the PN detector or to increase the statistics of a specific observation. For very bright sources (with 
fluxes of the order of $10^{-10}$~erg~cm$^{-2}$~s$^{-1}$ or more), only the PN data in small window mode have 
been analyzed.

The spectra were rebinned so that each energy bin contained a minimum of 20 counts and fit in the $0.4-10$~keV 
energy range for the PN detector and $0.5-10$~keV for MOS detectors, because of the uncertainties in the
calibration and cross-calibration at lower energies (cf. Kirsch 2005). The fluxes and luminosities were
calculated in the $0.4-10$~keV band by extrapolating the model spectrum with the command \texttt{extend} 
of \texttt{xspec}. The photon redistribution matrix and the related ancillary files were created appropriately with the 
\texttt{rmfgen} and \texttt{arfgen} tasks of \texttt{XMM-SAS}.

The data from the Optical Monitor (Mason et al. 2001) were also reprocessed with the latest version of \texttt{SAS}.

Through the paper, we report only the fits with reduced  $\chi^2$ less than 2 (${\tilde{\chi}^2}<2$) 
and we consider significant improvement in the fit with $Ftest>99$\%. In the case of an \emph{added} spectral 
component, we evaluated the improvement of the fit by the $\Delta \chi^2$ value (cf Protassov et al. 2002). All the quoted 
uncertainties in the parameters are at the $90$\% confidence level for 1 parameter ($\Delta \chi^2=2.71$), unless 
otherwise stated.

\section{Average spectra and comparisons with other catalogs}
Most of the sources analyzed here are blazars, except for two radiogalaxies Fanaroff-Riley Type I (FRI), 
that are thought to be blazar-like sources seen at large viewing angles (Urry \& Padovani 1995).
Blazars display featureless X-ray continuum, occasionally with hints of breaks, curvature, or, more seldom, a soft 
excess (e.g. Giommi et al. 2002, Donato et al. 2005, Perlman et al. 2005). Radiogalaxies can have a much more complex
environment at low energies (e.g. Evans et al. 2005, Grandi et al. 2005). However, since the main purpose of this
work is to study and compare the continuum properties, we decided to fit the X-ray spectra with a redshifted power-law model 
(\texttt{zpo} in \texttt{xspec}; see Table~\ref{tab:srcdata}) and a broken power law model (\texttt{bknpo} in \texttt{xspec};
see Table~\ref{tab:srcdata2}). The absorption can be fixed to the Galactic column density (Dickey \& Lockman 1990) or
fitted. Sometimes a more complex fit is necessary and is analyzed separately. More details on the fits are available
in Appendix A (tables and notes on the individual sources).

\begin{figure*}[!t]
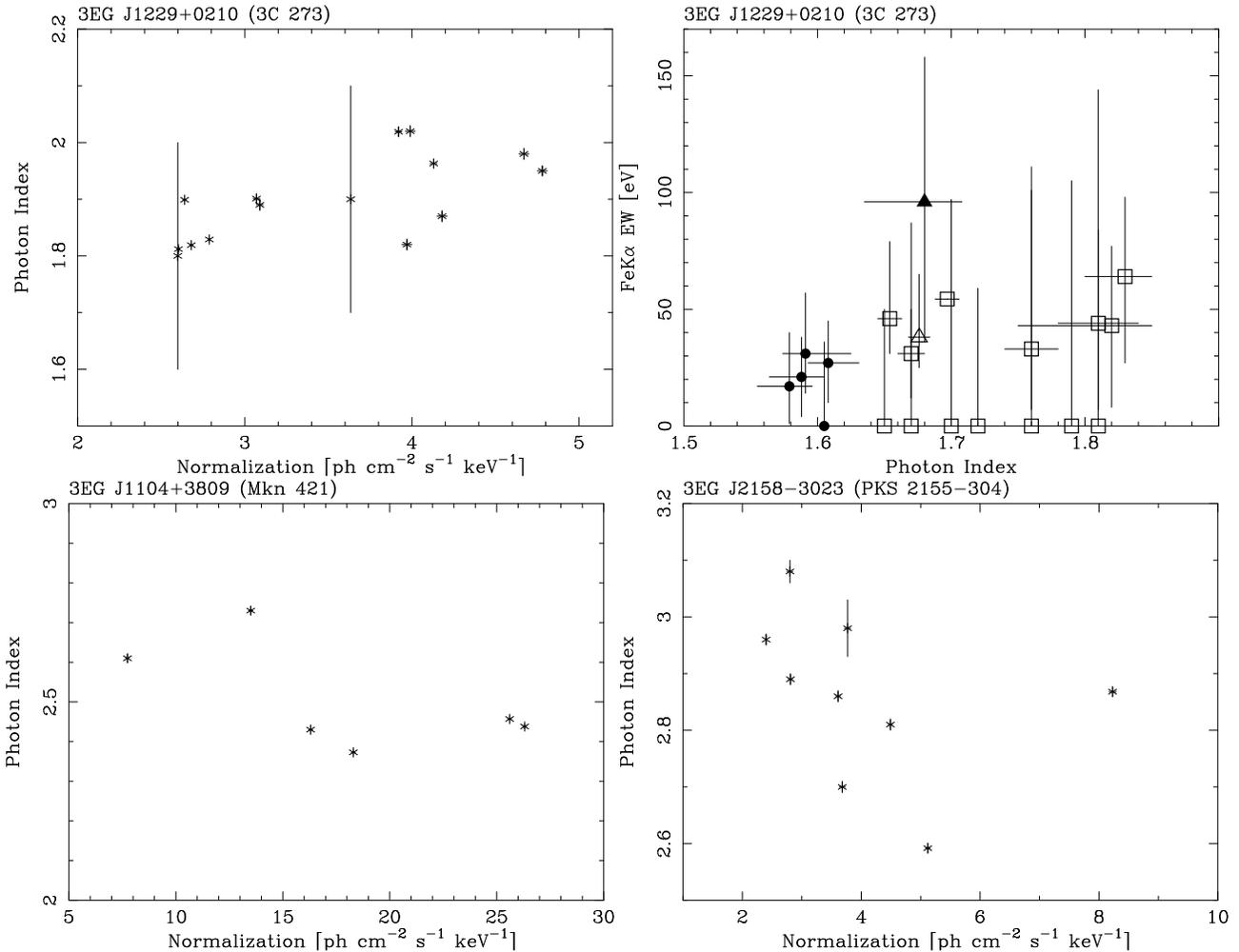

\begin{center}
\includegraphics[scale=0.35,angle=270]{4921_f1a.ps}
\includegraphics[scale=0.35,angle=270]{4921_f1b.ps}\\
\includegraphics[scale=0.35,angle=270]{4921_f1c.ps}
\includegraphics[scale=0.35,angle=270]{4921_f1d.ps}
\caption{Correlations: Photon index versus normalization for (\emph{top left}) 3C~$273$ (3EG~J$1229+0210$), 
(\emph{bottom left}) Mkn~$421$ (3EG~J$1104+3809$), and (\emph{bottom right}) PKS~$2155-304$ (3EG~J$2158-3023$).
(\emph{top right}) Equivalent width of the FeK$\alpha$ line versus photon index for 3C$273$. The open square 
indicates the \emph{XMM-Newton} data from the present work, while the filled circle represents the values from 
\emph{BeppoSAX} observations reported in Grandi \& Palumbo~(2004). The points of the simultaneous 
\emph{BeppoSAX}-\emph{XMM-Newton} observation (June 2001) are tagged with a triangle, filled for 
\emph{BeppoSAX} and open for \emph{XMM-Newton}.}
\label{correlaz}
\end{center}
\end{figure*}

Several of the $15$ sources analyzed here show significant variability, particularly on small time scales. Therefore, 
the comparison of average spectral parameters presented here (Table~\ref{tab:averages}) 
and the values available in the \emph{BeppoSAX} catalogues by Giommi et al. (2002) and Donato et al. (2005), can be 
considered a indicator of long term spectral variability, since \emph{XMM-Newton} observations refer to the
period $2000-2004$ and \emph{BeppoSAX} data have been collected in the period $1996-2002$. The spectra of the present 
work are generally best fitted with broken power law models (9/14 in 
Table~\ref{tab:averages} without PKS~$2155-304$, that is fitted with both models), compared to only 4/14 
sources in \emph{BeppoSAX} data (Table~\ref{tab:bepposax}, without 3C~273, that is fitted with both models). 
There is a possible important bias factor in \emph{BeppoSAX} fits. The Italian-Dutch 
satellite concentrators LECS and MECS have energy bands overlapping at $\approx 2$~keV, where most of the blazars 
have the break energy. Therefore, it is also possible that -- in some cases -- the intercalibration constant between the 
two detectors could have ``absorbed'' some spectral shape variations, thus leading to prefer the single power law model. This could be the
case of PKS~$1127-145$, that is known to have an intervening system along the line of sight, but the \emph{BeppoSAX} fit
does not require an additional absorption. 

In other cases, there are changes in the source state (e.g. AO~$0235+164$, that in the present observation
was found in outburst) or 3C~$273$, as already noted by Page et al. (2004). In the latter case, the blazar shows
in the \emph{XMM-Newton} data an increase of the break energy and a small softening of $\Gamma_2$, the photon 
index at $E>E_{\rm break}$. Also Cen~A shows a softer photon index and a flux lower by a factor of $4$ with respect to 
the \emph{BeppoSAX} data analyzed in Grandi et al. (2003). In the two HBL (Mkn~$421$ and PKS~$2155-304$), the broken power 
law model appears to be the simplification of a more complex curved model, as already noted by Brinkmann et al. (2001, 2003), 
Sembay et al. (2002), Ravasio et al. (2004). 

Comparing the parameters in Table~\ref{tab:averages} with the larger catalogs by Giommi et al. (2002), Donato et al. 
(2005), Evans et al. (2005), Grandi et al. (2005) containing also radio-loud AGN not detected by EGRET, there are no 
signs of differences between $\gamma-$ray loud and quiet AGN. 

The analysis of the three most intensively observed sources is in agreement with the typical behaviour of these types 
of sources. Even though they require a broken power law model, to have an overall view of 
the source behaviour, we correlate the photon index of the single power law model with its normalization at $1$~keV. 
This gives an idea of the behaviour of the spectral shape with flux variations. 

PKS~$2155-304$ shows a clear correlation (linear correlation coefficient $r=-0.76$ for $8$ observations and $p\leq 0.035$), with a hardening 
of the photon index with increasing flux (Fig.~\ref{correlaz}, \emph{bottom right}) if one point is not considered
(ObsID $0124930301$). While the behaviour of most 
points can be explained by considering a flux increase with a constant synchrotron peak (implying a hardening of the 
spectrum), the outlying point can be explained as due to a frequency shift of the synchrotron peak. For 
this ObsID, OM data (Table~\ref{tab:omdata}) suggest a high flux level in the bands U, B, and V, but, again, available 
magnitudes are fragmentary and it is not possible to search for more stringent correlations. 

The correlation in Mkn~$421$ is poor ($r=-0.60$ for 6 points, $p\leq 0.2$), although a general trend of 
spectral hardening with flux increasing can be noted (Fig.~\ref{correlaz}, \emph{bottom left}).

3C~$273$ shows instead a spectral softening with flux increasing (Fig.~\ref{correlaz}, \emph{top left}, $r=0.66$ with
15 observations, $p\leq 0.0077$), although there is a non negligible scatter of the points, suggesting that other processes 
are playing important roles in this sources. Indeed, it is known that this blazar also has a Seyfert-like component that can 
be detected (Grandi \& Palumbo 2004).
Fig.~\ref{correlaz} (\emph{top right}) shows the equivalent width of the FeK$\alpha$ line versus the photon index for both
\emph{BeppoSAX} observations (filled circles; data from Grandi \& Palumbo 2004) and the present work (open squares; 
for \emph{XMM-Newton} see Table~\ref{tab:ironlines}).
3C~$273$ was in a different state during the two satellites observations with only one point 
overlapping, when using the simultaneous \emph{BeppoSAX} and \emph{XMM-Newton} observation performed in June~$2001$ 
(ObsID~$0136550101$, triangles in Fig.~\ref{correlaz} \emph{top right}) to cross-calibrate the respective instruments (cf Molendi \& Sembay 2003). 
This strenghtens the validity of the other values obtained by the two satellites as indicating an effective change in the state of the source. The 
general trend can be understood, in the framework of the Grandi \& Palumbo (2004) results on 3C~$273$ and the more general 
picture on blazars outlined by Maraschi \& Tavecchio (2003), as a weakening of the jet component and an increase of the Seyfert-like part.
With respect to the \emph{BeppoSAX} observations ($1996-2002$), we noted in the \emph{XMM-Newton} observations 
($2000-2004$), an increase of the ``thermal'' component and a softening of the hard photon index
The former is indicated in the broken power law model by a shift to high energies of the break energy (see 
Table~\ref{tab:averages}: the average value for \emph{XMM-Newton} is $E_{\rm break}=1.44\pm 0.08$~keV compared to 
the \emph{BeppoSAX} value of $0.9\pm 0.3$~keV); in the blackbody plus power law model (Table~\ref{tab:srcdata3}) this is 
indicated by an increase of the temperature (from the average value of $54_{-4}^{+6}$~eV measured by \emph{BeppoSAX} to 
$143\pm 6$~eV derived from the data analyzed in the present work). This behaviour of the continuum is accompanied by 
an increase of the equivalent width of the (broad) iron emission line either at $6.4$ or $6.7$~keV (Fig.~\ref{correlaz}, \emph{top right}).
According to the results in Table~\ref{tab:ironlines}, the major improvements in the fit occur with the detection of the 
(broad) iron line centered at $6.4$ or $6.7$~keV, but sometimes there is no detection at all. If we consider that the 
detection and the energy centroid are related to different degrees of ionization (see, for example, the review by 
Reynolds \& Nowak 2003), the interpretation of the data still favours the hypothesis of an increase of the accretion around
the SMBH. A more detailed spectral analysis is required to better assess the state variation of 3C~$273$, but this is
outside the scope of the present work. However, the general trend outlined here is also confirmed by the radio data 
reported, for example, by Ter\"asranta et al. (2005) with observations at $22$ and $37$~GHz: there is a decreasing of activity 
from $1996$, with an outburst in $2003$, but smaller than in $1998-1999$. 

Another interesting case in the present sample is AO~$0235+164$, that displays the typical characteristics
of blazars in outburst. This should be compared with the negative detection of variability reported in the $2004$ 
observations (Raiteri et al. 2005).
A more detailed analysis of all the \emph{XMM-Newton} data sets (both public and private) is available in
Raiteri et al. (2006). In the present observation, the source displays a shift of the synchrotron peak (see the 
discussion in Section~6), with a hint of periodicity of $7829$~s (significance $4.5\sigma$), but with a low quality factor, 
because of a limited number of cycles ($2.15$). An inspection of the lightcurve suggests that this periodicity is transient: 
periodic flares appear during the activity phase of the source, while X-ray observations of the source in quiescence 
have detected constant flux and, obviously, no periodicities (Raiteri et al., 2006).

\section{Intervening absorption}
Torres et al. (2003) have suggested that gravitational microlensing can boost the $\gamma-$ray flux. Therefore,
we searched in the present sample for any presence of intervening systems of any type.

Damped Lyman$-\alpha$ (DLA) systems (see Wolfe et al. 2005 for a review) are present along the line of sight 
toward AO~$0235+164$ and PKS~$1127-145$. In the case of AO~$0235+164$ ($z=0.94$, Cohen et al. 1987), 
the intervening system is placed along the line of sight at $z=0.524$ (see, e.g., Raiteri et al. 2005 for 
a more recent discussion on this). This intervening system has been measured by \emph{ROSAT} and \emph{ASCA} 
obtaining a value of $N_{\rm H}^{z}=(2.8\pm 0.4)\times 10^{21}$~cm$^{-2}$ (Madejski et al., 1996). In addition,
Raiteri et al. (2005) measured a value of $N_{\rm H}^{z}=(2.4\pm 0.2)\times 10^{21}$~cm$^{-2}$ with an \emph{XMM-Newton}
observation performed on January $18$, $2004$. The present spectrum in the $0.4-10$~keV energy 
band for PN and the $0.5-10$~keV band for the two MOS detectors is the best fit with a broken power law model absorbed by the 
Galactic column together with one at redshift $z=0.524$ with $N_{\rm H}^{z}=(2.56\pm 0.05)\times 10^{21}$~cm$^{-2}$. 
The low energy photon index is $\Gamma_1=2.50\pm 0.01$ and the high-energy one is 
$\Gamma_2=2.05\pm 0.03$, with the break at $3.08_{-0.12}^{+0.09}$~keV, for a ${\tilde{\chi}^2}=1.10$ and $801$ dof (see
also Appendix A). 

PKS~$1127-145$ is another quasar with an additional absorber at $z=0.321$ along the line of sight, probably due to two 
late-type galaxies (Bergeron \& Boiss\'e 1991, Lane et al. 1998). The best fit model is the broken power law, with the 
$z=0.321$ absorption column of $(1.2_{-0.1}^{+0.2})\times 10^{21}$~cm$^{-2}$, the Galactic column, 
and no intrinsic absorption. \emph{Chandra} observations reported the same absorption, but the continuum is best fitted 
with a single power law with $\Gamma=1.19\pm 0.02$ (Bechtold et al. 2001). 

The absorption in X-rays, in both the above cases, is lower than that measured from optical observations, but it can 
be due to different metallicity ($Z<Z_{\odot}$) or to the presence of dust with a ratio different from that in the Galaxy, 
that is common in DLA systems (Pei et al. 1991, Pettini et al. 1994). However, what is important in these systems is 
the presence of galaxies along the line of sight that can cause gravitational lensing effects, which in turn enhance
the flux of the background blazar.

Another blazar (S5~$0836+710$) presents an intervening system at $z=0.914$, but optical observations by 
Stickel \& K\"uhr (1993) indicated the presence of Mg II $\lambda\lambda 2796,2803$, which do not qualify this system
as damped Lyman$-\alpha$. The blazar spectrum, already quite hard, presents a spectral flattening at low energies. 
This type of flattening has been observed in other blazars at high redshift ($z>2$; see, e.g., PMN~J$0525-3343$ 
in Fabian et al. 2001 or on RBS~$315$ in Piconcelli \& Guainazzi 2005) and a few hypotheses have been invoked, like
instrinsic absorption or intrinsic spectral properties. The best fit for S5~$0836+710$ is that with the additional 
absorption at the redshift of the quasar and no improvements are obtained by adding an absorber at the redshift of
the intervening system ($z=0.914$). Therefore, it appears that this system is not responsible for the additional 
absorption or -- at least -- the present data do not allow us to separate the different components, if any. Radio 
observations by Hutchison et al. (2001) show that near the blazar core the polarization is very low with respect to typical 
values in other quasars. The authors explained this by the presence of an external environment surrounding the jet 
(a cocoon?). Such a plasma cocoon could act like the ``warm absorber'' of Seyferts and
explain the spectral flattening at low energies. Yet another hypothesis is a low-energy cut-off in the relativistic 
electron distribution at $\gamma_{\rm min}$, which would yield a flattening below 
$\approx \gamma^{2}_{\rm min}\Gamma^{2}\nu_{\rm ext}$, where $\nu_{\rm ext}$ is the peak frequency of the seed photon
radiation field. These hypotheses are discussed in, e.g., Fabian et al. (2001).

Two more sources in the present sample show evidence of intrinsic absorption and are the two radiogalaxies 
(Cen~A and NGC~$6251$). This is known to be due to the environment in the radiogalaxies (cf Evans et al. 2005, 
Grandi et al. 2005). 

The last case of intervening systems is PKS~$1830-211$ ($z=2.507$), that is gravitationally lensed by a galaxy
at $z=0.886$ found by Wiklind \& Combes (1996) through infrared observations of hydrocarbon absorption lines. 
The fit with a redshifted power law model absorbed by the Galactic column and the intervening system at $z=0.886$, 
gives results consistent with previous X-ray analyses by Mathur \& Nair (1997), Oshima et al. (2001), and De Rosa et al. 
(2005). The absorption due to the intervening system averaged over the three observations is 
$N_{\rm H}^{z}=(2.3\pm 0.1)\times 10^{22}$~cm$^{-2}$ and the photon index is $\Gamma=1.14\pm 0.02$. However, some 
residuals at low energy are present and an improvement in the fit (with $\Delta \chi^2=30.9$,~$103$, and $16.6$ for 
a decrease of two degrees of freedom, for the three observations respectively) can be obtained by adding a thermal 
plasma model (\texttt{mekal}) at the redshift of the blazar, with solar abundances and temperature $kT=0.39\pm 0.06$~keV. 
The absorption of the intervening system is slightly greater ($N_{\rm H}^{z}=(3.0\pm 0.4)\times 10^{22}$~cm$^{-2}$) and 
the photon index a little steeper ($\Gamma=1.21\pm 0.04$). This thermal plasma can be the ``warm absorber'' suggested by 
Fabian et al. (2001) to explain the X-ray deficit at low energies in high redshift blazars. Interestingly, the statistical 
best fit of these three \emph{XMM-Newton} observations is obtained with a broken power law model and lower absorption 
(Table~\ref{tab:srcdata2}), thus suggesting that the low X-ray deficit could be due to something intrisic to the 
electron distribution.

\begin{figure}[!t]
\begin{center}
\includegraphics[scale=0.8]{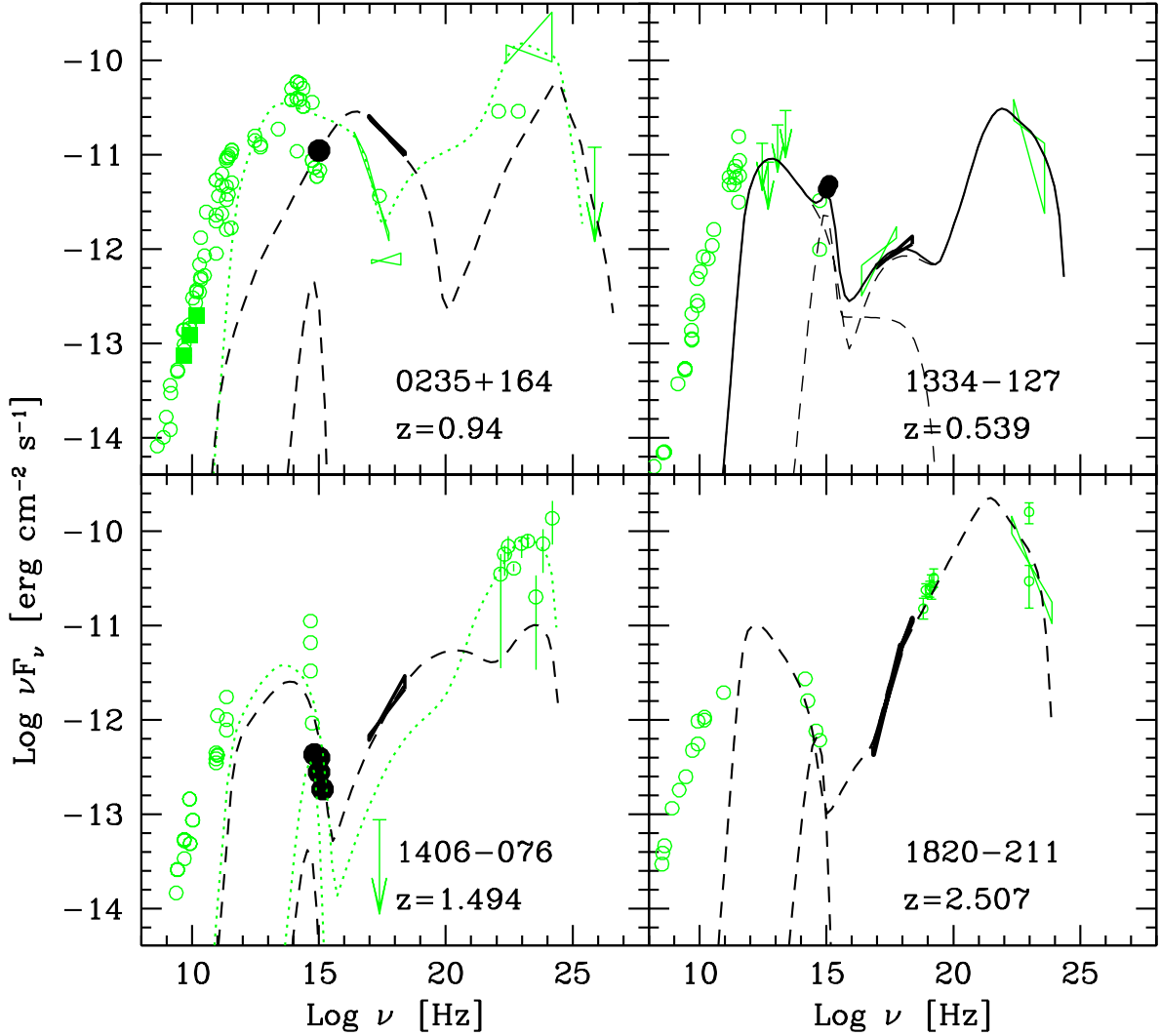}
\caption{SED of newly modelled blazars: (\emph{top, left}) AO~$0235+164$; (\emph{top, right}) PKS~$1334-127$; 
(\emph{bottom, left}) PKS~$1406-076$; (\emph{bottom, right}) PKS~$1830-211$, not corrected for the gravitational 
lensing effects. The black line is a single--zone synchrotron inverse Compton model using the input 
parameters listed in Table~\ref{param}. In all the SED, the \emph{XMM--Newton} data are indicated with black filled symbols, 
while the remaining symbols and lines (grey) refer to archival data. The bump at $\sim 10^{14-15}$ Hz is the assumed disk spectrum.}
\label{seds}
\end{center}
\end{figure}

\section{X-ray spectral line features}
The spectra of the AGN in the present sample sometimes show some features, but these are generally within $2\sigma$ 
deviations from the best fit model. For more significative detections (e.g. in PKS~$2155-304$, known to be due to 
warm-hot intergalactic medium, see Cagnoni et al. 2004), we bypassed the problem with a proper selection of the 
energy band, since the study of these features is outside the aims of the present work. No evidence ($>3\sigma$) 
of features linked to physical characteristic of any cosmic source is found in the present data set, except for the iron 
line complex of Cen~A (see Table~\ref{tab:srcdata4} and the note on this source in Appendix A) and some detections in 3C~$273$ 
(Table~\ref{tab:ironlines}). The latter is not always evident, although a forced fit with a broad iron line both neutral 
and ionized can give sometimes a non negligible improvement in the $\chi^2$. The implications have been already discussed in the Sect.~3.

\section{Spectral Energy Distributions and the blazar sequence}
SED have been constructed and modelled to study the multiwavelength emission over a broad energy range.
All but two (PKS~$1334-127$ and PKS~$1830-211$) of the blazars in the present sample have been studied in 
detail by Ghisellini et al. (1998), Tagliaferri et al. (2000), Ghisellini, Celotti \& Costamante  (2002). 
The radiogalaxies have been studied in Chiaberge et al. (2001, 2003), Guainazzi et al. (2003), 
Foschini et al. (2005), and Ghisellini et al. (2005). We refer to these papers and to the references therein. 
The SED with \emph{XMM-Newton} data are reported in Fig.~\ref{sed1} and \ref{sed2},
together with the model of synchrotron and inverse Compton radiation (including self-Compton and external Compton) 
in a homogeneous region applied to the data. The results obtained do not change dramatically with respect to the 
above mentioned works.

There are however four cases, namely AO~$0235+164$, PKS~$1334-127$, PKS~$1406-76$ and PKS~$1830-211$, which are 
worth investigating further. AO~$0235+164$ shows a clear shift in its peak frequency, as expected 
during flaring activities of blazars: the present \emph{XMM-Newton} observation
was performed when the source was in outburst (see Sect.~$3$). PKS~$1334-127$ has been associated with an EGRET 
source only in the Third Catalog (Hartman et al. 1999), and therefore is missing in Ghisellini et al. (1998). 
PKS~$1406-076$ was never detected in X-rays and PKS~$1830-211$ is a gravitationally lensed system, also missing
in Ghisellini et al. (1998). 

\begin{table}[!t]
\begin{center}
\caption{Parameters for the models of AO~$0235+164$, PKS~$1334-127$, PKS~$1406-076$, and PKS~$1830-211$: 
above the dividing line are listed the input parameters used to model the SED according to the finite injection 
time synchrotron--inverse Compton model of Ghisellini, Celotti \& Costamante (2002). Below the dividing line we 
list the output parameters for the four fitted sources. See the text for more details.} 
\begin{tabular}{llllll}
\hline
Parameter            & AO~$0235+164$ & PKS~$1334-127$ & PKS~$1406-076$  & PKS~$1830-211$ & Units \\
\hline
$R$                  & $4.5$         & $3.5$          & $2.0$           & $2.0$           & $10^{16}$~cm \\
$\Delta R$           & $2.8$         & $3.5$          & $2.0$           & $30$            & $10^{15}$~cm \\
$L^\prime_{\rm inj}$ & $7.5$         & $3.0$          & $3.0$           & $7.0$           & $10^{43}$~erg~s$^{-1}$  \\
$\gamma_{\rm break}$     & $2.0\times 10^{4}$ & $2.5\times 10^{2}$ & $1.0\times 10^{2}$ & $1.4\times 10^{2}$  & \\
$\gamma_{\rm max}$   & $6.0\times 10^{5}$ & $4.0\times 10^{3}$ & $5.0\times 10^{3}$ & $1.6\times 10^{3}$  & \\
$s$                  & $2.55$        & $2.7$          & $2.2$           & $2.8$           & \\
$B$                  & $1.3$         & $3.7$          & $1.0$           & $3.8$           & Gauss \\
$\Gamma$             & $16$          & $10$           & $10$            & $17$   		  & \\
$\theta$             & $3$           & $6$            & $3.6$           & $3.5$           & degree \\
$\delta$             & $18.2$        & $9.6$          & $14.3$          & $16.4$          &  \\
$L_{BLR}$            & $3.0$         & $4.5$          & $0.8$           & $12$            & $10^{44}$~erg~s$^{-1}$  \\
$R_{BLR}$            & $7.0$         & $3.5$          & $5.0$           & $3.5$           & $10^{17}$~cm \\
\hline
$\gamma_{\rm peak}$  & $2.0\times 10^{4}$ & $2.5\times 10^{2}$ & $1.6\times 10^{3}$       & $1.4\times 10^{2}$   & \\
$U_B$                & $0.067$       & $0.54$         & $0.04$          & $0.57$          & erg~cm$^{-3}$\\
$U^\prime_{\rm r}(\gamma_{\rm peak})$ & $0.042$ & $1.8$ & $0.25$        & $10.87$         & erg~cm$^{-3}$\\
$L_B$                & $3.28$        & $6.26$         & $0.15$          & $14$            & $10^{45}$~erg~s$^{-1}$\\
$L_{\rm e}$          & $1.05$        & $1.78$         & $8.8$           & $1.0$           & $10^{45}$~erg~s$^{-1}$\\
$L_{\rm p}$          & $7.16$        & $3.15$         & $25$            & $26.5$          & $10^{46}$~erg~s$^{-1}$\\
$L_{\rm rad}$        & $4.588$       & $3.59$         & $3.2$           & $20.8$          & $10^{45}$~erg~s$^{-1}$\\
\hline
\end{tabular}
\label{param}
\end{center}
\end{table}

We therefore applied the same model used in Ghisellini, Celotti \& Costamante (2002) in order to find out the 
physical parameters of the four sources. The main assumptions of the model can be summarized as follows:
the geometry of the source is a cylinder -- except for PKS~$1830-211$, that is analyzed later -- of radius $R$ and 
length, in the comoving frame, $\Delta R'=R/\Gamma$, where $\Gamma$ is the bulk Lorentz factor; $\theta$ is the 
viewing angle, $\delta$ the Doppler factor, and $B$ the magnetic field. The radiating particle distribution is 
assumed to be $N(\gamma)\propto \gamma^{-s}$, where the value of $s$ depends on the value of $\gamma$ of the 
injected particles. The injected power in the comoving frame is $L^\prime_{\rm inj}$.
The external seed photon field has a dimension $R_{BLR}$ and luminosity $L_{BLR}$. It is assumed 
to mainly originate in a Broad Line Region or any other external source and it is calculated as a fraction
of the disk luminosity (generally 10\%). The magnetic and radiative energy densities are indicated with $U_B$ and 
$U_{\rm r}$, respectively. Electron, proton and radiation powers are represented by $L_{\rm e}$, $L_{\rm p}$, 
and $L_{\rm rad}$, respectively.

Fig.~\ref{seds} show the model with the SED, while in Table~\ref{param} we list the input and output parameters for 
the model. Only AO~$0235+164$ displays significant variability in the model parameters with respect to Ghisellini 
et al. (1998), but it continues to fulfill the requirements of the blazar sequence. PKS~$1334-127$ behaves as a 
typical FSRQ. 

The X-ray emission of PKS~$1406-076$ is modelled as due to the inverse Compton emission from synchrotron 
seed photons, although this source is a FSRQ. An external source of seed photons is needed to 
generate $\gamma-$rays in the EGRET energy band (not simultaneous data) and there is an apparent anti-correlation 
between X-ray and optical/UV emission: indeed, this first X-ray detection is simultaneous with low optical flux, while archival data 
(not simultaneous) report higher optical flux and only an upper limit in X-rays (for more details, see the note on this 
source in the Appendix). This does not allow us to claim an anti-correlation.

In the case of PKS~$1830-211$, there are some problems that should be taken into account: the magnification 
effects of the gravitational lensing are still uncertain (cf Oshima et al. 2001 and Courbin et al. 2002) and so as
the absorption, both due to the Galactic column (the source is apparently located at low Galactic latitude, with 
Galactic coordinates $l=12^{\circ}.16$ and $b=-5^{\circ}.71$) and to the intervening system. Therefore, instead of
making hypotheses about the quantity and quality of corrections to be applied, we decided to analyze the observed SED 
without any correction. This means that the anomalies in the parameters of PKS~$1830-211$ (Table~\ref{param}) 
reflect the uncertainties in the magnification and the absorption. For example, the needed $L_{BLR}$ necessary for 
the external Compton contribution was calculated as $30$\% of the disk luminosity, while for the three other blazars 
a value of $10$\% was taken. A proper dereddening could result in the needed optical flux, without invoking an 
increase of percentage of the disk luminosity. 

\begin{figure}[!t]
\begin{center}
\includegraphics[scale=0.4]{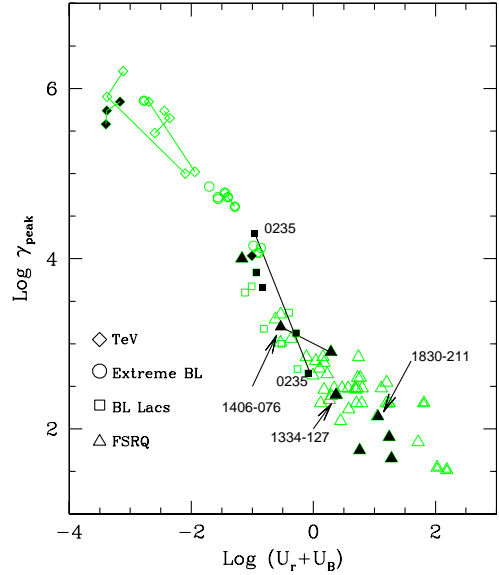}
\caption{Updated blazar sequence, adapted from Ghisellini, Celotti \& Costamante~(2002), with the addition 
of the new TeV BL Lacs. We mark with black filled symbols the sources analyzed in this paper, emphasizing
AO~$0235+164$ (in quiescence and during the outburst presented in this work), PKS~$1334-127$, PKS~$1406-076$ (the fit
without the X-ray detection and with the present data), and PKS~$1830-211$.}
\label{ubs}
\end{center}
\end{figure}

The location of the sources analyzed here in the blazar sequence can be seen in Fig.~\ref{ubs}, where we show 
$\gamma_{\rm peak}$, the Lorentz factor of the electrons radiating mostly at the peak of the SED, versus 
$U_B+U^\prime_{\rm rad}$, the radiation plus magnetic energy density in the comoving frame.
Fig.~\ref{ubs} has been updated by adding 3 new BL Lac recently detected in the TeV range (see Aharonian et 
al. 2005a, 2005b). These new TeV BL Lacs, namely 1ES~$1101-232$ (cf also Wolter et al. 2000), 
PKS~$2005-489$, and H~$2356-309$, lie in the ``high energy branch" defined by the BL Lacs previously 
detected in the TeV band. Interestingly, PKS~$1406-076$ moved toward the region of the BL Lac region, while the previous
modeling -- with only an upper limit in X-rays -- placed this FSRQ in the region typical of these sources. 

\begin{table}[!t]
\begin{center}
\caption{Parameters useful to understand $\gamma-$ray loudness. Columns: (1) Source name;
(2) beaming factor $\delta$; (3) observed flux in the $0.4-10$~keV energy band [erg~cm$^{-2}$~s$^{-1}$];
(4) intrinsic luminosity in the $0.4-10$~keV energy band [erg~s$^{-1}$]; (5) Confidence of the EGRET
detection (high $> 95$\%; low $< 95$\%).} 
\begin{tabular}{lcccc}
\hline
Source         & $\delta$  & $F$                & $L$  & Conf.\\
(1)            & (2)       & (3)                & (4)  & (5)\\
\hline
3C~$273$       & $6.5-7$   & $\approx 10^{-10}$ & $\approx 10^{46}$ & high\\
NGC~$6251$     & $3.2-3.8$ & $\approx 10^{-12}$ & $\approx 10^{43}$ & low\\
PKS~$0521-365$ & $1.4-3$   & $\approx 10^{-11}$ & $\approx 10^{42}$ & low\\
Cen~A          & $1.2-1.6$ & $\approx 10^{-10}$ & $\approx 10^{41}$ & high\\
\hline
\end{tabular}
\label{beaming}
\end{center}
\end{table}

The presence of two radiogalaxies in the present sample suggests some hints about the paradigm of the unification
of radio-loud AGN (Urry \& Padovani 1995), of which the $\gamma-$ray loud AGN are the subclass analyzed in this work.
This paradigm and the theories on the blazar evolution (see B\"ottcher \& Dermer 2002, Cavaliere \& D'Elia 2002, Maraschi
\& Tavecchio 2003) find analogies between BL Lac and FRI on one side and FSRQ and FRII on the other. BL Lac and FRI are
evolved AGN, with low emission from the environment around the SMBH, while FSRQ and FRII are instead young sources
with a rich environment.

With reference to the $\gamma-$ray propagation from the source to the observer, one of the most important factor
is the beaming factor $\delta$, that allow high energy photons to escape from the source without disappearing in pair
production. From the SED of the $\gamma-$ray loud AGN reported in Ghisellini et al. (1998, 2005), Chiaberge et al. (2001,
2003), Foschini et al. (2005), complemented and confirmed by this work, we see that the $\delta$ values 
for almost all the EGRET detected AGN are above $10$, with only a few exceptions. 3C~$273$ with $\delta=6.5-7$ and 
Cen~A with $\delta=1.2-1.6$ are also the only AGN with $\delta < 10$ detected by EGRET with high confidence. 
PKS~$0521-365$ ($\delta=1.4-3$) and NGC~$6251$ ($\delta=3.2-3.8$) have low confidence identifications, that should be
confirmed. The two $\delta$ of the low confidence detections are in between 3C~$273$ and Cen~A,
which are instead detected at high confidence level. Therefore, there should be other reasons to explain the EGRET detections.
Indeed, three of the four above sources have also the lowest intrinsic luminosities in the present sample (and also 
among the whole EGRET sample), but not the fluxes (cf Table~\ref{beaming}). This implies 
that the present definition of $\gamma-$ray loudness -- that it is defined here simply as the detection at 
$E > 100$~MeV -- is still strongly biased by the instrument sensitivity or by the distance 
of the source.

Moreover, there is another key point still missing in this picture: $\gamma-$ray detection of FRII 
radiogalaxies, that are still completely missing to date even in the list of hypothetical associations.

\section{Final remarks}
A small sample of AGN $\gamma-$ray loud (i.e. detected by the EGRET instrument on board \emph{CGRO}) observed by 
\emph{XMM-Newton} has been analyzed in a homogeneous way and presented here. 
The sample is composed of 15 AGN divided into 7 FSRQ, 4 LBL, 2 HBL, and 2 FRI radiogalaxies. 
All the data were taken from the public archive of \emph{XMM-Newton}: 46 pointings were analyzed, of which 
30 are of three sources only (3C~$273$, Mkn~$421$, PKS~$2155-304$). 

Despite these limitations, some useful inferences can be made. Indeed, with \emph{XMM-Newton} it is possible
to perform simultaneous X-ray and optical/UV observations, that can be particularly useful in blazars to place
reliable constraints on the synctrotron and inverse Compton peaks in the SED. 

The main findings can be summarized as follows: all the blazars obey the sequence suggested 
by Ghisellini et al. (1998) and Fossati et al. (1998). The only X-ray features found in the present sample 
are the emission lines of the iron complex in Cen~A and in 3C~$273$. In the case of Cen~A, the iron line at 
$6.4$~keV ($\sigma \lesssim 80$~eV) is known to be due to the transmission of radiation along the dust lane 
warped around Cen~A, while in 3C~$273$, the broad iron line can be associated with the Seyfert-like component.

The comparison with \emph{BeppoSAX} data show a preference of the broken power law model over the single 
power law; the latter was often the best fit in the \emph{BeppoSAX} catalog, suggesting a long term change in the sources.
The changes in the spectra of 3C~$273$ appear to be genuinely due to a variation in the state of the source, as well 
as in the case of AO~$0235-164$, observed during an outburst. 

Four sources show intervening systems along the line of sight, but only one case is known to be gravitationally lensed. 
In the remaining three cases, it is not clear if the intervening galaxies can generate gravitational effects strong enough 
to enhance the $\gamma-$ray loudness. See Torres et al. (2003) for a discussion on this topic.

The SED compiled in the present sample confirm the model parameters already found in previous studies and increase the
number of modelled sources. However, PKS~$1406-076$ shows some particular features and deserves further investigation.
Four sources appear to be the key to understand the transition (with respect to $\gamma-$ray loudness) from blazars to 
radiogalaxies, namely 3C~$273$, PKS~$0521-365$, NGC~$6251$, and Cen~A. The first two are FSRQ 
with the largest jet viewing angle, which in turn means the lowest $\delta$ among the blazars. The latter two are the (only) 
radiogalaxies detected by EGRET. Further and more detailed studies on these four sources could give important contributions
to the comprehension of the mechanisms acting to generate $\gamma-$rays, of the unification models of AGN, and to the 
improvement of the resolution of the extragalactic $\gamma-$ray background.

\begin{acknowledgements}
LF thanks G.G.C. Palumbo, P. Grandi and M. Dadina for useful discussions and S.R. Rosen of the OM
Team for useful hints in the OM data analysis. We thank also the referee, X. Barcons, for useful comments, that helped
to improve the manuscript. 
This research has made use of the NASA's Astrophysics Data System Abstract Service and of the NASA/IPAC Extragalactic 
Database (NED), which is operated by the Jet Propulsion Laboratory, California Institute of Technology, under contract 
with the National Aeronautics and Space Administration. This work was partly supported by the European 
Community's Human Potential Programme under contract HPRN-CT-2002-00321 and by the Italian Space Agency (ASI).
\end{acknowledgements}

\Online
\appendix
\section{Notes on individual sources, tables and SED}
\begin{figure*}[!ht]
\begin{center}
\includegraphics[scale=0.8]{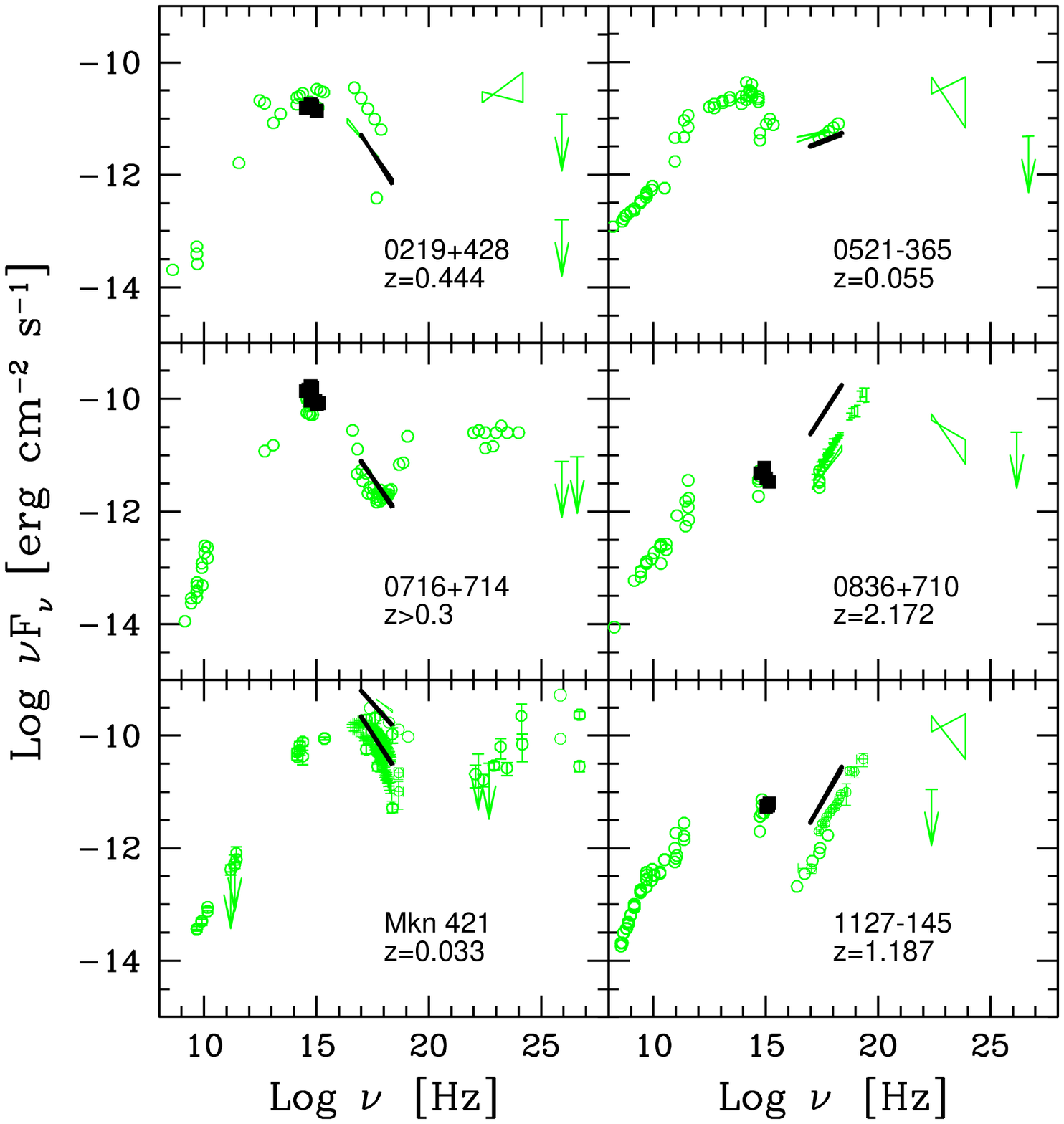}
\caption{SED of the sources studied in this paper. We compare the \emph{XMM-Newton} data (filled black symbols) to the 
available archival data in all bands.}
\label{sed1}
\end{center}
\end{figure*}

\begin{figure*}[!ht]
\begin{center}
\includegraphics[scale=0.8]{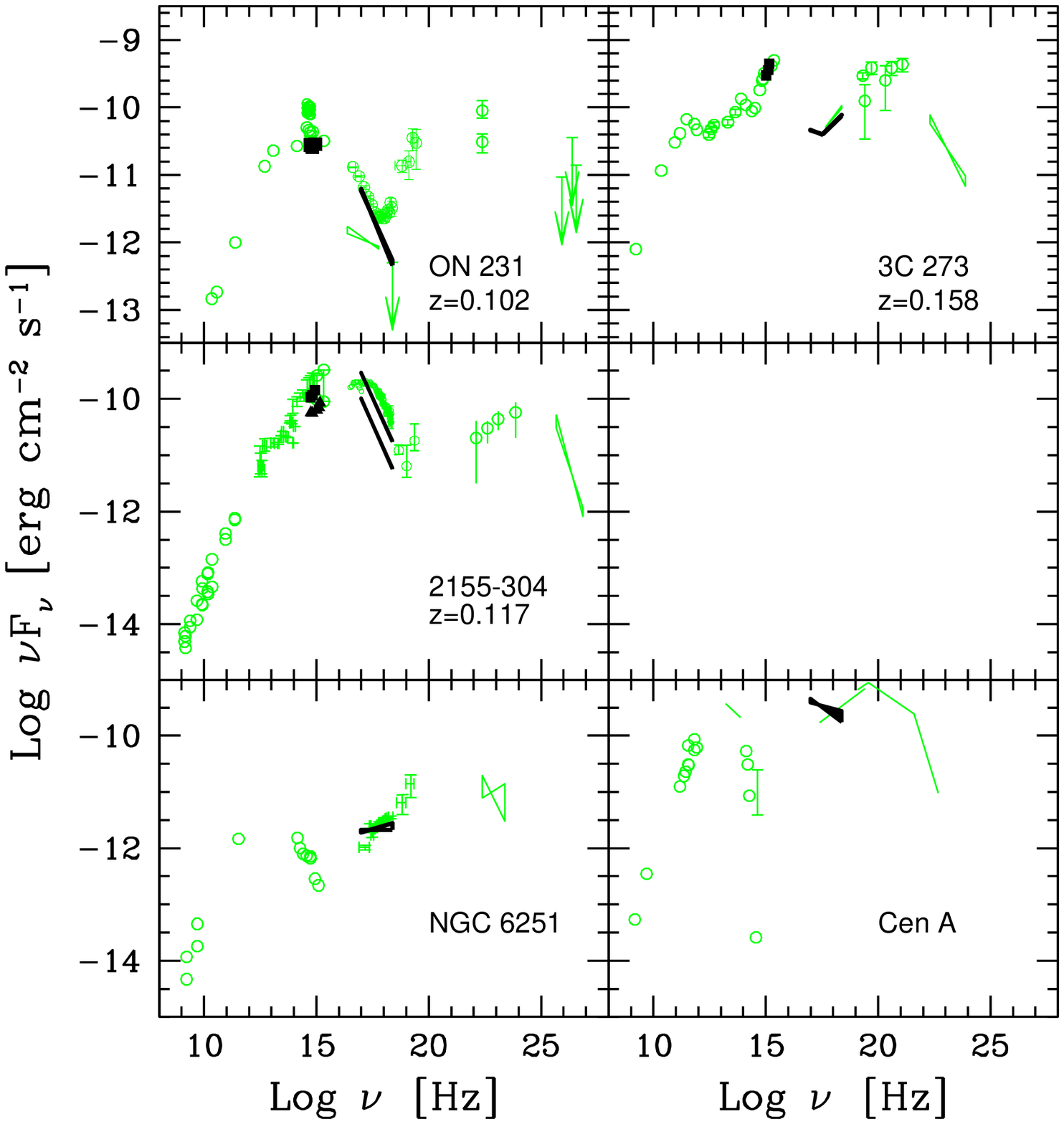}
\caption{SED of the sources studied in this paper. We compare the \emph{XMM-Newton} data (filled black symbols) to the 
available archival data in all bands.}
\label{sed2}
\end{center}
\end{figure*}

We report in this Appendix the tables with the Observation log (Table~\ref{tab:log}), the fit with the simple
power law model (Table~\ref{tab:srcdata}), the broken power law model (Table~\ref{tab:srcdata2}),
the additional fits for 3C~$273$ (Table~\ref{tab:srcdata3} and \ref{tab:ironlines}) and Cen~A (Table~\ref{tab:srcdata4})
and the magnitudes with different filters of the Optical Monitor (Table~\ref{tab:omdata}). Some notes on the
individual sources and the SED of the sources not reported in Sect.~6 (Fig.~\ref{sed1} and \ref{sed2}) complete 
this Appendix.

\begin{table*}[!ht]
\caption{\emph{XMM-Newton} Observation Log. Columns: (1) Source name; (2) Observation Identifier;
(3) Date of the observation [DD-MM-YYYY]; (4,5,6) Observing mode of MOS1, MOS2, and PN, respectively
[FF: Full Frame; SW: Small Window; TIMING: timing mode, without imaging] with the effective exposure 
time [ks]; (7) Angular distance from the boresight [arcsec].}
\centering
\begin{tabular}{llllllr}
\hline
Name          & ObsID        & Date         & MOS1   & MOS2   & PN       & Position\\
(1)          & (2)          & (3)          & (4)    & (5)    & (6)      & (7)\\
\hline
$0219+428$ & $0002970201$ & $05-02-2002$ & FF(10)   & FF(10)   & FF(11.0) & $366$ \\
AO~$0235+164$ & $0110990101$ & $10-02-2002$ & FF(19)   & FF(19)   & FF(15.0) & $77$  \\
PKS~$0521-365$ & $0065760201$ & $09-10-2002$ & SW(31.3) & FF(31.3) & FF(27.1) & $65$ \\
S5~$0716+714$ & $0150495601$ & $04-04-2004$ & SW(31)   & FF(31)   & TIMING   & $69$\\
S5~$0836+710$ & $0112620101$ & $12-04-2001$ & SW(23.8) & SW(23.8) & FF(24.6) & $14$  \\
Mkn~$421^{\mathrm{a}}$ & $0099280201$ & $01-11-2000$ & SW       & SW       & SW(24.2)  & $6.5$  \\
{}           & $0099280301$ & $13-11-2000$ & SW       & SW       & SW(25.6)  &  $7.1$ \\
{}           & $0099280401$ & $14-11-2000$ & SW       & SW       & SW(23.4)  &  $119.7$ \\
{}           & $0136540101$ & $08-05-2001$ & SW       & SW       & SW(25.7)  &  $4.7$ \\
{}           & $0158970101$ & $01-06-2003$ & TIMING   & SW       & SW(25.3)  &  $6.1$ \\
{}           & $0162960101$ & $10-12-2003$ & SW       & SW       & SW(17.4)  &  $6.4$ \\
PKS~$1127-145$ & $0112850201$ & $01-07-2002$ & FF(13.9) & FF(13.7) & FF(10.7) & $68$  \\
ON~$231$     & $0104860501$ & $26-06-2002$ & FF(33.1) & FF(32.6) & FF(26.7) & $472$ \\
3C~$273^{\mathrm{a}}$ & $0126700301$ & $13-06-2000$ & SW       & SW       & SW(39.7) & $10$  \\
{}           & $0126700601$ & $15-06-2000$ & SW       & SW       & SW(20.8) & $9$ \\
{}           & $0126700701$ & $15-06-2000$ & SW       & SW       & SW(21.0) & $8$ \\
{}           & $0126700801$ & $17-06-2000$ & SW       & SW       & SW(42.5) & $9$ \\
{}           & $0136550101$ & $13-06-2001$ & SW       & SW       & SW(62.0) & $5$ \\
{}           & $0112770101$ & $16-12-2001$ & TIMING   & SW       & SW(3.5)  & $54.8$\\
{}           & $0112770201$ & $22-12-2001$ & TIMING   & SW       & SW(3.5)  & $54.5$\\
{}           & $0112770601$ & $07-07-2002$ & TIMING   & SW       & SW(3.5)  & $78.7$ \\
{}           & $0112770801$ & $17-12-2002$ & TIMING   & SW       & SW(3.5)  & $55.0$ \\
{}           & $0136550501$ & $05-01-2003$ & SW       & SW       & SW(6.0)  & $5$ \\
{}           & $0112770701$ & $05-01-2003$ & TIMING   & SW       & SW(3.5)  & $58.8$ \\
{}           & $0112771001$ & $18-06-2003$ & TIMING   & SW       & SW(3.9)  & $79.1$ \\
{}           & $0112770501$ & $08-07-2003$ & TIMING   & SW       & SW(5.6)  & $78.0$\\
{}           & $0112771101$ & $14-12-2003$ & TIMING   & SW       & SW(5.9)  & $57.2$ \\
{}           & $0136550801$ & $30-06-2004$ & SW       & SW       & SW(13.9) & $3$ \\
Cen~A        & $0093650201$ & $02-02-2001$ & FF(22.8) & FF(22.8) & FF(16.8) & $9$\\
{}           & $0093650301$ & $06-02-2002$ & FF(13.2) & FF$^{\mathrm{b}}$ & FF(7.9)  & $16$\\
PKS~$1334-127$ & $0147670201$ & $31-01-2003$ & FF(13.5) & FF(13.5) & FF(10.9) & $68$ \\
PKS~$1406-076$ & $0151590101$ & $05-07-2003$ & FF(7.8)  & FF(10.0) & FF(7.3)  & $68$ \\
{}           & $0151590201$ & $10-08-2003$ & FF(5.6)  & FF(5.8)  & FF(3.6)  & $69$ \\
NGC~$6251$   & $0056340201$ & $26-03-2002$ & FF(18.0)  & FF(18.0)  & FF(8.0)  & $69$ \\
PKS~$1830-211$ & $0204580201$ & $10-03-2004$ & FF(7.0)  & FF(7.6)  & FF(2.8)  & $67$\\
{}           & $0204580301$ & $24-03-2004$ & FF(31.0) & FF(31.0) & FF(27.0)  & $68$\\
{}           & $0204580401$ & $05-04-2004$ & FF(18.5)  & FF(18.5)  & FF(13.0)  & $68$\\
PKS~$2155-304^{\mathrm{a}}$ & $0124930201$ & $31-05-2000$ & TIMING   & SW       & SW(41.6) & $9$  \\
{}           & $0080940101$ & $19-11-2000$ & TIMING   & SW       & SW(40.2) & $10$ \\
{}           & $0080940301$ & $20-11-2000$ & TIMING   & SW       & SW(40.8) & $16$ \\
{}           & $0124930301$ & $30-11-2001$ & SW       & SW       & SW(31.2) & $15$ \\
{}           & $0124930501$ & $24-05-2002$ & SW       & SW       & SW(22.3) & $6$  \\
{}           & $0124930601$ & $29-11-2002$ & SW       & SW       & SW(39.8) & $2$  \\
{}           & $0158960101$ & $23-11-2003$ & SW       & SW       & SW(18.7) & $5$  \\
{}           & $0158960901$ & $22-11-2004$ & SW       & SW       & SW(20.0) & $6$  \\
{}           & $0158961001$ & $23-11-2004$ & SW       & SW       & SW(28.0) & $6$  \\
\hline
\end{tabular}
\begin{list}{}{}
\item[$^{\mathrm{a}}$] Only PN data have been analyzed, because the high flux of the source caused strong pile-up.
\item[$^{\mathrm{b}}$] MOS2 not used, because of a series of bad pixels in the source PSF.
\end{list}
\label{tab:log}
\end{table*}

\vskip 12pt
\emph{3EG~J0222+4253 (0219+428, 3C~66A):} The counterpart of this EGRET source is the BL Lac $0219+428$, 
although Kuiper et al.~(2000) have proposed that the flux below $300$~MeV is significantly 
contaminated by the nearby pulsar PSR~J$0218+4232$. 3C~$66$A was extensively monitored
from radio to very high $\gamma-$rays during $2003-2004$ (B\"ottcher et al. 2005) and no 
significant X-ray variability was detected.
The present \emph{XMM-Newton} data set was affected by a high background, but it is 
possible to use more than $70$\% of the observation. 
The fit of the spectrum of this BL Lac object with an absorbed power law model is 
acceptable (see Table~\ref{tab:srcdata}), although there are residuals for energies 
greater than $4$~keV. The broken power law model provides the best fit 
with a confidence $>99.99$\% with a f-test.
This observation has been studied by Croston et al.~(2003) and the fit with a 
simple power law absorbed by the Galactic $N_{\rm H}$ is consistent with
the present results. However, significant differences are present in the fit with the
broken power law model: this can be understood by taking into account that in the present
analysis a more conservative selection of events has been used (for example, this can be clearly
inferred by comparing the PN exposures: $11$~ks in the present analysis vs $15$~ks in the 
analysis of Croston et al., 2003).

\vskip 12pt

\begin{table*}[!ht]
\caption{Fit results with single powerl law model. Columns: (1) Source name; 
(2) absorbing column density  [$10^{20}$~cm$^{-2}$]; 
(3) photon index of the power law; 
(4) Normalization of the power law model [$10^{-2}$~ph~cm$^{-2}$~s$^{-1}$~keV$^{-1}$ at $1$~keV]; 
(5) Reduced $\chi^2$ and degrees of freedom;
(6) observed flux in the $0.4-10$~keV energy band [$10^{-11}$~erg~cm$^{-2}$~s$^{-1}$];
(7) intrinsic luminosity in the $0.4-10$ keV energy band rest frame [$10^{45}$ erg s$^{-1}$].
The uncertainties in the parameter estimates are at the $90\%$ confidence limits for 1 parameter. 
For $N_{\mathrm{H}}=\mathrm{Gal}$, it means that the absorption column has been fixed to the Galactic value.}
\centering
\begin{tabular}{lcccccc}
\hline
Name         & $N_{\mathrm{H}}$  & $\Gamma$         & $A$              & $\tilde{\chi}^2$/dof   & $F$      & $L$\\
(1)         & (2)                & (3)               & (4)              & (5)              & (6)              & (7) \\
\hline
$0219+428$ & Gal.              & $2.60\pm 0.04$   & $0.183\pm 0.007$ & $1.29/293$     & $0.21$  & $2.6$\\
AO~$0235+164$ & Gal.              & $2.28\pm 0.02$   & $1.22\pm 0.04$   & $0.99/618$     & $0.96$  & $66$\\
PKS~$0521-365$ & Gal.              & $1.847\pm 0.009$ & $0.231\pm 0.002$ & $1.15/1202$    & $1.2$   & $9.0\times 10^{-4}$\\
S5~$0716+714^{\rm a}$ & Gal.      & $2.58\pm 0.02$   & $0.294\pm 0.003$ & $1.74/419$     & $1.0$   & $>14.2$\\
S5~$0836+710^{\rm b}$ & $14\pm 3$ & $1.379\pm 0.007$ & $2.62\pm 0.04$   & $1.07/2272$    & $4.8$   & $902$\\
Mkn~$421^{\rm c}$ & $2.3\pm 0.3$ & $2.61\pm 0.01$ & $7.74\pm 0.09$  & $0.97/1026$    & $26.0$  & $0.59$\\
{}           & $3.5\pm 0.2$      & $2.438\pm 0.006$ & $26.3\pm 0.1$    & $1.20/1491$    & $90.6$  & $2.12$\\
{}           & $3.1\pm 0.2$      & $2.457\pm 0.006$ & $25.6\pm 0.1$    & $1.29/1430$    & $88.8$  & $2.05$\\
{}           & $3.1\pm 0.2$      & $2.373\pm 0.007$ & $18.3\pm 0.1$    & $1.12/1447$    & $66.1$  & $1.51$\\
{}           & $4.6\pm 0.2$      & $2.73\pm 0.01$   & $13.5\pm 0.1$    & $1.16/1178$    & $40.6$  & $1.02$\\
{}           & $3.7\pm 0.2$      & $2.43\pm 0.01$   & $16.3\pm 0.1$    & $1.03/1307$    & $56.0$  & $1.31$\\
PKS~$1127-145^{\rm d}$ & $12_{-1}^{+2}$ & $1.31\pm 0.02$   & $0.34\pm 0.01$   & $1.01/890$     & $1.1$   & $59.2$\\
ON~$231$     & $2.5\pm 0.6$      & $2.77\pm 0.04$   & $0.186\pm 0.005$ & $1.07/653$     & $0.50$  & $0.16$\\
3C~$273$     & Gal.              & $1.829\pm 0.004$ & $2.788\pm 0.007$ & $1.82/1598$    & $12.2$  & $8.4$\\
{}           & Gal.              & $1.819\pm 0.005$ & $2.68\pm 0.01$   & $1.71/1401$    & $11.8$  & $8.2$\\
{}           & Gal.              & $1.812\pm 0.005$ & $2.603\pm 0.009$ & $1.53/1394$    & $11.5$  & $8.0$\\
{}           & Gal.              & $1.8\pm 0.2$     & $2.599\pm 0.004$ & $2.08/1647$    & $11.5$  & $8.0$\\
{}           & Gal.              & $1.9\pm 0.2$     & $3.634\pm 0.004$ & $4.22/1711$    & $14.5$  & $10.2$\\
{}           & Gal.              & $1.87\pm 0.01$   & $4.18\pm 0.03$   & $1.24/802$     & $17.4$  & $12.2$\\
{}           & Gal.              & $1.82\pm 0.01$   & $3.97\pm 0.03$   & $1.15/806$     & $17.5$  & $12.1$\\
{}           & Gal.              & $1.89\pm 0.01$   & $3.09\pm 0.02$   & $1.08/680$     & $12.7$  & $8.90$\\
{}           & Gal.              & $1.98\pm 0.01$   & $4.67\pm 0.03$   & $1.21/790$     & $17.7$  & $12.6$\\
{}           & Gal.              & $2.019\pm 0.009$ & $3.92\pm 0.02$   & $1.37/899$     & $14.3$  & $10.3$\\
{}           & Gal.              & $2.02\pm 0.01$   & $3.99\pm 0.03$   & $1.23/711$     & $14.5$  & $10.4$\\
{}           & Gal.              & $1.950\pm 0.009$ & $4.78\pm 0.03$   & $1.27/826$     & $18.5$  & $13.2$\\
{}           & Gal.              & $1.963\pm 0.008$ & $4.13\pm 0.02$   & $1.12/908$     & $15.8$  & $11.2$\\
{}           & Gal.              & $1.901\pm 0.009$ & $3.07\pm 0.02$   & $1.33/877$     & $12.5$  & $8.77$\\
{}           & Gal.              & $1.899\pm 0.006$ & $2.64\pm 0.01$   & $1.32/1160$    & $10.7$  & $7.54$\\
Cen~A$^{\rm e}$ & $1400\pm 200$ & $2.2\pm 0.1$ & $22_{-5}^{+7}$   & $0.97/668$     & $16.9$  & $3.0\times 10^{-4}$\\
{}           & $1800\pm 300$         & $2.3\pm 0.2$ & $29_{-9}^{+14}$  & $0.97/340$     & $18.5$  & $3.3\times 10^{-4}$\\
PKS~$1334-127$ & $6.7\pm 0.9$      & $1.80\pm 0.04$   & $0.171\pm 0.009$ & $0.93/609$     & $0.43$  & $5.1$\\
PKS~$1406-076$ & Gal.              & $1.58\pm 0.07$   & $0.060_{-0.006}^{+0.007}$& $0.79/90$  & $0.10$ & $10.5$\\
{}           & Gal.              & $1.6\pm 0.1$     & $0.056_{-0.007}^{0.009}$ & $0.94/49$  & $0.094$ & $9.78$\\
NGC~$6251$ & $11.6_{-0.7}^{+0.8}$  & $1.94\pm 0.03$   & $0.134\pm 0.004$ & $1.04/812$     & $0.577$ & $9.6\times 10^{-3}$\\
PKS~$1830-211^{\rm d}$ & $350\pm 50$ & $1.17_{-0.05}^{+0.03}$ & $0.62_{-0.08}^{+0.07}$ & $0.90/600$ & $1.42$ & $322$\\
{}           & $240\pm 10$         & $1.14\pm 0.02$   & $0.54\pm 0.03$   & $1.04/1662$    & $1.34$  & $290$\\
{}           & $220\pm 10$         & $1.13\pm 0.03$	  & $0.47\pm 0.03$   & $1.04/1168$    & $1.22$  & $261$\\
PKS~$2155-304^{\rm c}$ & $1.6\pm 0.2$ & $2.592\pm 0.009$ & $5.12\pm 0.04$ & $1.02/1266$   & $14.4$  & $5.71$\\
{}           & $2.2\pm 0.2$      & $2.81\pm 0.01$   & $4.49\pm 0.04$   & $1.00/1067$    & $11.6$  & $4.87$\\
{}           & $1.9\pm 0.3$      & $2.86\pm 0.01$   & $3.61\pm 0.04$   & $1.03/991$     & $9.32$  & $3.90$\\
{}           & $3.2\pm 0.2$      & $2.868\pm 0.009$ & $8.23\pm 0.06$   & $1.01/1131$    & $20.2$  & $8.89$\\
{}           & $1.4\pm 0.3$      & $2.70\pm 0.01$   & $3.68\pm 0.05$   & $0.99/886$     & $10.0$  & $4.02$\\
{}           & $3.0\pm 0.3$      & $2.89\pm 0.01$   & $2.81\pm 0.04$   & $0.87/880$     & $6.90$  & $3.03$\\
{}           & Gal.              & $2.96\pm 0.01$   & $2.40\pm 0.01$   & $0.95/614$     & $6.17$  & $2.60$\\
{}           & $2.5\pm 0.5$      & $3.08\pm 0.02$   & $2.80\pm 0.05$   & $0.88/625$     & $6.83$  & $3.04$\\
{}           & $2.9\pm 0.4$      & $2.98\pm 0.02$   & $3.77\pm 0.05$   & $1.10/831$     & $9.14$  & $4.07$\\
\hline
\end{tabular}
\begin{list}{}{}
\item[$^{\mathrm{a}}$] Only MOS1+MOS2, since the PN was set in TIMING. Lower limit of luminosity calculated for $z=0.5$.
\item[$^{\mathrm{b}}$] Additional absorber placed at the redshift of the source (\texttt{wa*zwa(zpo)} model in \texttt{xspec}).
\item[$^{\mathrm{c}}$] Fit in the $0.6-10$~keV energy range and extrapolation to $0.4$~keV for flux and luminosity calculations.
\item[$^{\mathrm{d}}$] Additional redshifted absorber (\texttt{wa*zwa(zpo)} model in \texttt{xspec}) along the line of sight.
\item[$^{\mathrm{e}}$] Fit in the $4-10$~keV energy range, without $6-8$~keV energy band, because of complex features 
in the low energy part and in the iron emission line complex.
\end{list}
\label{tab:srcdata}
\end{table*}

\begin{table*}[!ht]
\caption{Fit results with the broken power law model. Columns: (1) Source name; (2) absorbing column density  
[$10^{20}$~cm$^{-2}$]; (3) low energy photon index of the power law; (4) high energy photon index; (5) Break energy [keV]; 
(6) Normalization of the power law model [$10^{-2}$~ph~cm$^{-2}$~s$^{-1}$~keV$^{-1}$ at $1$~keV]; 
(7) Reduced $\chi^2$ and degrees of freedom;
(8) observed flux in the $0.4-10$~keV energy band [$10^{-11}$~erg~cm$^{-2}$~s$^{-1}$];
(9) intrinsic luminosity in the $0.4-10$ keV energy band rest frame [$10^{45}$ erg s$^{-1}$];
(10) Ftest probability [\%]; fit with probability below $90$\% are not reported.
The uncertainties in the parameter estimates are at the $90\%$ confidence limits for 1 parameter. 
For $N_{\mathrm{H}}=\mathrm{Gal}$, it means that the absorption column has been fixed to the Galactic value.}
\centering
\begin{tabular}{lccccccccc}
\hline
Name         & $N_{\mathrm{H}}$   & $\Gamma_1$             & $\Gamma_2$             & $E_{\rm break}$      & $A$                     & $\tilde{\chi}^2$/dof   & $F$      & $L$      & $Ftest$\\
(1)         & (2)                & (3)                    & (4)                    & (5)                  & (6)                     & (7)              & (8)      & (9)      & (10)\\
\hline
$0219+428$ & Gal.              & $2.91_{-0.08}^{+0.12}$ & $2.23_{-0.09}^{+0.10}$ & $1.3\pm 0.2$         & $0.067\pm 0.003$        & $1.05/291$       & $0.24$  & $3.0$     & $>99.99$\\
AO~$0235+164$ & Gal.              & $2.33\pm 0.04$         & $2.1\pm 0.1$           & $3.3_{-0.5}^{+0.7}$  & $0.276\pm 0.006$        & $0.96/616$       & $0.99$  & $69$      & $>99.99$\\
PKS~$0521-365$ & Gal.              & $1.95\pm 0.03$         & $1.74\pm 0.03$         & $1.5_{-0.2}^{+0.3}$  & $0.227\pm 0.003$        & $1.05/1200$      & $1.3$   & $9.4\times 10^{-4}$ & $99.97$\\
S5~$0716+714^{\rm a}$ & Gal.      & $2.70\pm 0.02$         & $1.98_{-0.09}^{+0.08}$ & $2.3_{-0.1}^{+0.2}$  & $0.293\pm 0.003$        & $1.06/417$       & $1.1$   & $>15.7$    & $>99.99$\\
S5~$0836+710$ & Gal.              & $1.25\pm 0.03$         & $1.375\pm 0.007$       & $1.01_{-0.08}^{+0.15}$& $0.529\pm 0.006$       & $1.07/2271$      & $4.8$   & $860$     & -\\
Mkn~$421^{\rm b}$ & Gal.      & $2.573\pm 0.008$       & $2.79_{-0.08}^{+0.17}$ & $4.2_{-0.7}^{+1.0}$  & $6.97\pm 0.02$          & $0.96/1025$      & $26.1$  & $0.578$   & $99.97$\\
{}           & Gal.              & $2.340\pm 0.004$       & $2.73\pm 0.04$         & $4.0\pm 0.2$         & $22.9\pm 0.4$           & $1.06/1490$      & $91.3$  & $2.00$    & $>99.99$\\
{}           & Gal.           & $2.378_{-0.005}^{+0.004}$ & $2.96_{-0.10}^{+0.08}$ & $4.7_{-0.3}^{+0.2}$  & $22.6\pm 0.4$           & $1.02/1429$      & $89.1$  & $1.95$    & $>99.99$\\
{}           & Gal.              & $2.312\pm 0.007$       & $2.68_{-0.05}^{+0.07}$ & $4.4\pm 0.3$         & $16.4\pm 0.7$           & $0.99/1446$      & $66.7$  & $1.45$    & $>99.99$\\
{}           & Gal.              & $2.59_{-0.02}^{+0.01}$ & $2.81\pm 0.02$         & $2.2_{-0.2}^{+0.1}$  & $11.5_{-0.8}^{+0.7}$    & $1.08/1177$      & $41.8$  & $0.929$   & $>99.99$\\
{}           & Gal.              & $2.300\pm 0.007$       & $2.51\pm 0.02$         & $2.2\pm 0.1$         & $14.1\pm 0.4$           & $0.99/1306$      & $56.3$  & $1.23$    & $>99.99$\\
PKS~$1127-145^{\rm c}$ & $12_{-1}^{+2}$ & $1.40_{-0.05}^{+0.08}$ & $1.22\pm 0.06$    & $2.7_{-0.8}^{+1.0}$  & $0.127_{-0.004}^{+0.006}$ & $1.00/888$       & $1.2$   & $61.0$    & $99.58$\\
ON~$231$ & Gal.              & $2.74\pm 0.02$         & $2.3_{-0.4}^{+0.3}$    & $4.2_{-1.1}^{+0.9}$  & $0.140\pm 0.002$        & $1.07/652$       & $0.51$  & $0.16$    & $94.50$\\
3C~$273$ & Gal.              & $1.943\pm 0.009$       & $1.697\pm 0.009$       & $1.48\pm 0.06$       & $2.073\pm 0.008$        & $1.13/1596$      & $12.7$  & $8.7$     & $>99.99$\\
{}           & Gal.              & $1.97\pm 0.01$         & $1.65\pm 0.01$         & $1.43\pm 0.08$       & $1.97\pm 0.01$          & $1.03/1399$      & $12.5$  & $8.6$     & $>99.99$\\
{}           & Gal.              & $1.96\pm 0.01$         & $1.67\pm 0.01$         & $1.35\pm 0.08$       & $1.91\pm 0.01$          & $1.00/1392$      & $12.1$  & $8.3$     & $>99.99$\\
{}           & Gal.              & $1.92\pm 0.01$         & $1.654\pm 0.009$       & $1.37\pm 0.05$       & $1.911\pm 0.008$        & $0.97/1645$      & $12.2$  & $8.4$     & $>99.99$\\
{}           & Gal.              & $2.121\pm 0.007$       & $1.676\pm 0.008$       & $1.45\pm 0.03$       & $2.604\pm 0.009$        & $1.10/1709$      & $15.7$  & $11.0$     & $>99.99$\\
{}           & Gal.              & $2.00\pm 0.02$         & $1.72\pm 0.02$         & $1.4_{-0.1}^{+0.2}$  & $3.06\pm 0.03$          & $1.00/800$       & $18.4$  & $12.7$     & $>99.99$\\
{}           & Gal.              & $1.94\pm 0.03$         & $1.67\pm 0.03$         & $1.5\pm 0.2$         & $2.95\pm 0.03$          & $0.95/804$       & $18.4$  & $12.6$     & $>99.99$\\
{}           & Gal.              & $1.97\pm 0.03$         & $1.79_{-0.04}^{+0.03}$ & $1.4_{-0.2}^{+0.4}$  & $2.29\pm 0.03$          & $1.01/678$       & $13.1$  & $9.09$     & $>99.99$\\
{}           & Gal.              & $2.09\pm 0.02$         & $1.81\pm 0.03$         & $1.6\pm 0.1$         & $3.39\pm 0.03$          & $0.97/788$       & $18.6$  & $13.2$     & $>99.99$\\
{}           & Gal.              & $2.16\pm 0.02$         & $1.83_{-0.03}^{+0.02}$ & $1.4_{-0.1}^{+0.2}$  & $2.80\pm 0.03$          & $0.99/897$       & $15.1$  & $10.8$     & $>99.99$\\
{}           & Gal.              & $2.13\pm 0.02$         & $1.81_{-0.05}^{+0.04}$ & $1.7_{-0.2}^{+0.3}$  & $2.88\pm 0.03$          & $0.96/709$       & $15.4$  & $10.9$     & $>99.99$\\
{}           & Gal.              & $2.05\pm 0.02$         & $1.76\pm 0.03$         & $1.8\pm 0.2$         & $3.51\pm 0.03$          & $1.02/824$       & $19.5$  & $13.7$     & $>99.99$\\
{}           & Gal.              & $2.02\pm 0.02$         & $1.82_{-0.07}^{+0.03}$ & $1.9_{-0.3}^{+0.6}$  & $3.05\pm 0.02$          & $1.00/906$       & $16.4$  & $11.6$     & $>99.99$\\
{}           & Gal.              & $2.02\pm 0.02$         & $1.70\pm 0.03$         & $1.7_{-0.1}^{+0.2}$  & $2.26\pm 0.02$          & $1.02/875$       & $13.2$  & $9.20$     & $>99.99$\\
{}           & Gal.              & $2.00_{-0.01}^{+0.02}$ & $1.76\pm 0.02$         & $1.5_{-0.1}^{+0.2}$  & $1.94\pm 0.01$          & $1.03/1158$      & $11.2$  & $7.8$      & $>99.99$\\
Cen~A        & {}                & {}                     & {}                     & {}                   & {}                      & {}               & {}      & {}         & -\\
{}           & {}                & {}                     & {}                     & {}                   & {}                      & {}               & {}      & {}         & -\\
PKS~$1334-127$ & Gal.              & $1.57_{-0.24}^{+0.08}$ & $1.78\pm 0.03$         & $1.0\pm 0.3$         & $0.077_{-0.003}^{+0.012}$ & $0.92/608$     & $0.43$  & $4.8$      & $92.31$\\
PKS~$1406-076$ & {}                & {}                     & {}                     & {}                   & {}                      & {}               & {} & {}              & $81.60$\\
{}           & {}                & {}                     & {}                     & {}                   & {}                      & {}               & {} & {}              & $56.30$\\
NGC~$6251$ & $14\pm 1$         & $2.11_{-0.06}^{+0.08}$ & $1.78\pm 0.07$         & $2.5_{-0.4}^{+0.3}$  & $0.141\pm 0.004$        & $1.00/810$       & $0.59$  & $0.010$    & $>99.99$\\
PKS~$1830-211$ & $62_{-6}^{+5}$    & $0.93_{-0.10}^{+0.08}$ & $1.42_{-0.05}^{+0.10}$ & $3.6_{-0.5}^{+0.4}$  & $0.11\pm 0.01$          & $0.83/598$       & $1.38$  & $270$      & $>99.99$\\
{}           & $62\pm 3$         & $0.89_{-0.09}^{+0.06}$ & $1.29\pm 0.04$         & $3.1\pm 0.3$         & $0.103_{-0.08}^{+0.06}$ & $0.97/1660$      & $1.31$  & $243$      & $>99.99$\\
{}           & $64\pm 3$         & $1.06\pm 0.04$         & $1.41_{-0.09}^{+0.12}$ & $4.7_{-0.4}^{+0.5}$  & $0.108\pm 0.06$         & $1.00/1166$      & $1.20$  & $248$      & $>99.99$\\
PKS~$2155-304^{\rm b}$ & Gal.      & $2.588\pm 0.005$       & $2.73_{-0.08}^{+0.18}$ & $4.4_{-1.1}^{+1.4}$   & $3.861\pm 0.008$       & $1.01/1265$      & $14.3$  & $5.70$     & $>99.99$\\
{}           & Gal.           & $2.773_{-0.006}^{+0.007}$ & $2.88_{-0.03}^{+0.06}$ & $2.8_{-0.4}^{+0.7}$   & $3.246\pm 0.007$       & $0.98/1066$      & $11.6$  & $4.74$     & $>99.99$\\
{}           & Gal.              & $2.845\pm 0.006$       & $3.0\pm 0.1$           & $4.3_{-1.2}^{+0.8}$   & $2.620\pm 0.007$       & $1.02/990$       & $9.31$  & $3.86$     & $99.67$\\
{}           & Gal.              & $2.783\pm 0.006$       & $2.96\pm 0.04$         & $2.4_{-0.2}^{+0.3}$   & $5.75\pm 0.01$         & $0.97/1130$      & $20.3$  & $8.35$     & $>99.99$\\
{}           & Gal.              & $2.75\pm 0.03$         & $2.70_{-0.02}^{+0.01}$ & $1.0_{-0.1}^{+0.5}$   & $2.74\pm 0.02$         & $0.99/885$       & $10.0$  & $4.10$     & $96.01$\\
{}           & {}                & {}                     & {}                     & {}                    & {}                     & {}               & {}      & {}         & -\\
{}           & {}                & {}                     & {}                     & {}                    & {}                     & {}               & {}      & {}         & $89.10$\\
{}           & {}                & {}                     & {}                     & {}                    & {}                     & {}               & {}      & {}         & -\\
{}           & {}                & {}                     & {}                     & {}                    & {}                     & {}               & {}      & {}         & $85.00$\\
\hline
\end{tabular}
\begin{list}{}{}
\item[$^{\mathrm{a}}$] Only MOS1+MOS2, since the PN was set in TIMING. Lower limit of luminosity calculated for $z=0.5$.
\item[$^{\mathrm{b}}$] Fit in the $0.6-10$~keV energy range and extrapolation to $0.4$~keV for flux and luminosity calculations.
\item[$^{\mathrm{c}}$] Additional redshifted absorber (\texttt{wa*zwa(bknpo)} model in \texttt{xspec}) along the line of sight.
\end{list}
\label{tab:srcdata2}
\end{table*}

\begin{table*}[!ht]
\caption{Additional fit results for 3C~$273$ (3EG~J$1229+0210$) with the model composed of a
black body plus a power law (\texttt{wa(zbb+zpo)}), absorbed with the Galactic column density. Columns: 
(1) Temperature [keV]; 
(2) Normalization of the black body model [$10^{-4}L_{39}/D_{10}^2$, where $L_{39}$ is the source luminosity in units
of $10^{39}$~erg/s and $D_{10}$ is the source distance in units of $10$~kpc]; ; 
(3) Photon index; 
(4) Normalization of the power law model [$10^{-2}$~ph~cm$^{-2}$~s$^{-1}$~keV$^{-1}$ at $1$~keV]; 
(5) Reduced $\chi^2$ and degrees of freedom;
(6) observed flux in the $0.4-10$~keV energy band [$10^{-11}$~erg~cm$^{-2}$~s$^{-1}$];
(7) intrinsic luminosity in the $0.4-10$ keV energy band rest frame [$10^{45}$ erg s$^{-1}$].
The uncertainties in the parameter estimates are at the $90\%$ confidence limits for 1 parameter.}
\centering
\begin{tabular}{ccccccc}
\hline
$kT$                      & $A_{\rm zbb}$          & $\Gamma$             & $A_{\rm pl}$            & $\tilde{\chi}^2$/dof   & $F$      & $L$      \\
(1)                & (2)                       & (3)                    & (4)                  & (5)                     & (6)     & (7) \\
\hline
$0.143\pm 0.004$          & $1.48\pm 0.07$         & $1.714\pm 0.005$     & $2.47\pm 0.02$          & $1.17/1596$      & $12.6$  & $8.7$     \\
$0.143\pm 0.004$          & $1.82_{-0.10}^{+0.09}$ & $1.668\pm 0.007$     & $2.29\pm 0.03$          & $1.06/1399$      & $12.5$  & $8.6$     \\
$0.138_{-0.004}^{+0.006}$ & $1.58_{-0.08}^{+0.11}$ & $1.685_{-0.008}^{+0.006}$ & $2.28_{-0.04}^{+0.02}$ & $1.01/1392$ & $12.1$  & $8.3$     \\
$0.137\pm 0.004$          & $1.76_{-0.10}^{+0.06}$ & $1.669\pm 0.009$     & $2.25\pm 0.02$          & $1.02/1645$      & $12.2$  & $8.3$     \\
$0.140\pm 0.002$          & $3.47\pm 0.07$         & $1.701\pm 0.005$     & $2.93\pm 0.02$          & $1.23/1709$      & $15.7$  & $10.9$    \\
$0.152_{-0.008}^{+0.007}$ & $2.8\pm 0.3$           & $1.72\pm 0.02$       & $3.54\pm 0.09$          & $0.99/800$       & $18.4$  & $12.7$     \\
$0.153\pm 0.009$          & $2.3\pm 0.3$           & $1.68\pm 0.02$       & $3.42\pm 0.09$          & $0.95/804$       & $18.3$  & $12.5$     \\
$0.16\pm 0.01$            & $1.4\pm 0.3$           & $1.79_{-0.02}^{+0.01}$ & $2.76\pm 0.08$        & $1.00/678$       & $13.1$  & $9.06$     \\
$0.149_{-0.008}^{+0.007}$ & $3.1_{-0.4}^{+0.3}$    & $1.82\pm 0.02$        & $4.0\pm 0.1$           & $0.98/788$       & $18.6$  & $13.1$     \\
$0.136\pm 0.006$          & $2.9\pm 0.2$           & $1.86\pm 0.01$       & $3.36\pm 0.07$          & $1.02/897$       & $15.0$  & $10.7$     \\
$0.148\pm 0.007$          & $3.1\pm 0.3$           & $1.84\pm 0.02$        & $3.33\pm 0.09$         & $0.95/709$       & $15.3$  & $10.9$     \\
$0.146_{-0.008}^{+0.007}$ & $3.0\pm 0.3$           & $1.80_{-0.02}^{+0.01}$& $4.13\pm 0.09$         & $1.04/824$       & $19.4$  & $13.6$     \\
$0.157\pm 0.008$          & $2.0\pm 0.3$           & $1.852_{-0.017}^{+0.009}$& $3.67\pm 0.08$      & $0.99/906$       & $16.3$  & $11.5$     \\
$0.146\pm 0.007$          & $2.1\pm 0.2$           & $1.74\pm 0.02$        & $2.61\pm 0.06$         & $1.04/875$       & $13.1$  & $9.13$     \\
$0.151\pm 0.005$          & $1.6\pm 0.1$           & $1.762\pm 0.008$     & $2.28\pm 0.04$          & $1.00/1158$      & $11.2$  & $7.8$     \\
\hline
\end{tabular}
\label{tab:srcdata3}
\end{table*}

\begin{table*}[!ht]
\caption{Detections and upper limits on the equivalent width of iron lines for 3C~$273$ (3EG~J$1229+0210$). 
The continuum is best fitted with the broken power law model with Galactic absorbtion reported in Table~\ref{tab:srcdata2}. 
Columns: 
(1) Equivalent width [eV] for $E=6.4$~keV and $\sigma=0.15$~keV;
(2) $\Delta \chi^2$ with respect to the best fit continuum (2 dof);
(3) Equivalent width [eV] for $E=6.4$~keV and $\sigma=0.5$~keV;
(4) $\Delta \chi^2$ with respect to the best fit continuum (2 dof);
(5) Equivalent width [eV] for $E=6.7$~keV and $\sigma=0.15$~keV;
(6) $\Delta \chi^2$ with respect to the best fit continuum (2 dof);
(7) Equivalent width [eV] for $E=6.7$~keV and $\sigma=0.5$~keV;
(8) $\Delta \chi^2$ with respect to the best fit continuum (2 dof);
The uncertainties in the parameters and upper limits estimate are at the $90\%$ confidence limits for 1 parameter.}
\centering
\begin{tabular}{cccccccc}
\hline
\multicolumn{4}{c}{$E=6.4$~keV} & \multicolumn{4}{c}{$E=6.7$~keV} \\
$\sigma=0.15$~keV  & $\Delta \chi^2$ & $\sigma=0.5$~keV & $\Delta \chi^2$ &  $\sigma=0.15$~keV & $\Delta \chi^2$ & $\sigma=0.5$~keV & $\Delta \chi^2$ \\
(1)                & (2)             & (3)              & (4)             & (5)                & (6)             & (7)              & (8)   \\
\hline
$29_{-12}^{+13}$   & $14.3$          & $54.5\pm 0.1$    & $20.2$          & $18.4\pm 0.1$      & $5.2$           & $39.3\pm 0.1$    & $7.7$\\
$<16$              & $-$             & $<50$            & $-$             & $<28$              & $-$             & $<53$            & $-$\\
$<19$              & $-$             & $<62$            & $-$             & $31\pm 19$         & $7.2$           & $<64$            & $-$\\
$20_{-12}^{+13}$   & $7.0$           & $46_{-15}^{+33}$ & $16.0$          & $21_{-14}^{+13}$   & $6.4$           & $62_{-36}^{+15}$ & $12.5$\\
$<20$              & $-$             & $38_{-13}^{+27}$ & $16.2$          & $22\pm 11$         & $10.8$          & $48_{-28}^{+13}$ & $12.0$\\
$<37$              & $-$             & $<59$            & $-$             & $<46$              & $-$             & $<83$            & $-$\\
$<56$              & $-$             & $<87$            & $-$             & $<30$              & $-$             & $<95$            & $-$\\
$<38$              & $-$             & $<105$           & $-$             & $<41$              & $-$             & $<113$           & $-$\\
$44_{-37}^{+40}$   & $3.3$           & $<117$           & $-$             & $<42$              & $-$             & $<133$           & $-$\\
$<62$              & $-$             & $80_{-67}^{+49}$ & $4.1$           & $64_{-37}^{+34}$   & $8.6$           & $110_{-83}^{+44}$& $5.8$\\
$<52$              & $-$             & $<144$           & $-$             & $<75$              & $-$             & $<156$            & $-$\\
$<42$              & $-$             & $<111$           & $-$             & $<60$              & $-$             & $<112$            & $-$\\
$<60$              & $-$             & $<106$           & $-$             & $43_{-35}^{+34}$   & $4.1$           & $<92$             & $-$\\
$<40$              & $-$             & $<97$            & $-$             & $<55$              & $-$             & $<125$            & $-$\\
$<40$              & $-$             & $33_{-26}^{+68}$ & $3.8$           & $<47$              & $-$             & $33_{-27}^{+74}$  & $3.7$\\
\hline
\end{tabular}
\label{tab:ironlines}
\end{table*}

\begin{table*}[!b]
\caption{(\emph{left}) Additional fit results for Cen~A (3EG~J$1324-4314$) with the model composed of an absorbed power law, plus one or two
gaussian emission lines. Fit performed in the $4-10$~keV energy band and then extrapolated to $0.4-10$~keV. 
The two columns refer to the two observations. The rows list the following parameters:
(1) Absorbing column density [$10^{20}$~cm$^{-2}$];
(2) Photon index of the power law model;
(3) Normalization of the power law model [$10^{-2}$~ph~cm$^{-2}$~s$^{-1}$~keV$^{-1}$ at $1$~keV]; 
(4) Energy of the emission line 1 [keV];
(5) Line width $\sigma$ 1 [keV];
(6) Flux of the emission line 1 [$10^{-4}$~ph~cm$^{-2}$~s$^{-1}$];
(7) Equivalent width of the emission line 1 [eV];
(8) Energy of the emission line 2 [keV];
(9) Line width $\sigma$ 2 [keV];
(10) Flux of the emission line 2 [$10^{-4}$~ph~cm$^{-2}$~s$^{-1}$];
(11) Equivalent width of the emission line 2 [eV];
(12) Reduced $\chi^2$ and degrees of freedom;
(13) observed flux in the $0.4-10$~keV energy band [$10^{-11}$~erg~cm$^{-2}$~s$^{-1}$];
(14) intrinsic luminosity in the $0.4-10$ keV energy band rest frame [$10^{45}$ erg s$^{-1}$].
The uncertainties in the parameter estimates are at the $90\%$ confidence limit for one parameter of interest. 
For $N_{\mathrm{H}}=\mathrm{Gal}$, it means that the absorption column has been fixed to the 
Galactic value. The luminosities were calculated using $d=3.84$~Mpc. 
(\emph{right}) EPIC spectrum in the $4-10$~keV energy band.}
\begin{minipage}{6cm}
\centering
\begin{tabular}{lcc}
\hline
{} & $0093650201$ & $0093650301$ \\
\hline
$N_{\mathrm{H}}$ & $1420_{-90}^{+40}$     & $1800\pm 200$\\
$\Gamma$         & $2.24_{-0.04}^{+0.07}$ & $2.3\pm 0.2$\\
$A_{\rm pl}$     & $22.1_{-1.2}^{+0.3}$   & $32_{-10}^{+13}$\\
$E_1$            & $6.41\pm 0.02$         & $6.44\pm 0.02$\\
$\sigma_1$       & $<0.074$               & $<0.065$ \\
$A_{\rm L1}$     & $2.6_{-0.6}^{+0.8}$    & $4.6_{-1.0}^{+1.1}$ \\
EqW$_1$          & $70_{-16}^{+22}$       & $107_{-23}^{+26}$\\
$E_2$            & $6.8_{-0.4}^{+0.3}$    & -\\
$\sigma_2$       & $0.76_{-0.25}^{+0.39}$ & -\\
$A_{\rm L2}$     & $4.3_{-2.3}^{+4.4}$    & -\\
EqW$_2$          & $133_{-71}^{+135}$     & -\\
$\tilde{\chi}^2$/dof &  $0.99/984$        & $0.96/498$\\
$F$              &  $17.4$                & $18.5$\\
$L$              &  $3.1\times 10^{-4}$  & $3.3\times 10^{-4}$\\
\hline
\end{tabular}
\end{minipage}
\begin{minipage}{12cm}
\centering
\includegraphics[scale=0.4,angle=270]{4921_tA6.ps}
\end{minipage}
\label{tab:srcdata4}
\end{table*}

\begin{table*}[!ht]
\caption{Optical properties of the AGN of the present catalog (from the Optical Monitor data). The data
refers to the magnitude averaged over the whole observation. 
Columns: 
(1) Source name; 
(2) V magnitude [$543$~nm]; 
(3) B magnitude [$450$~nm]; 
(4) U magnitude [$344$~nm]; 
(5) UVW1 magnitude [$291$~nm]; 
(6) UVM2 magnitude [$231$~nm];
(7) UVW2 magnitude [$212$~nm];
The uncertainties in the parameter estimates are at the $1\sigma$ level and also include systematics.}
\centering
\begin{tabular}{lcccccc}
\hline
Name          & V                 & B                 & U               & UVW1               & UVM2            & UVW2\\
(1)          & (2)               & (3)               & (4)             & (5)                & (6)             & (7) \\
\hline
$0219+428$    & {}                & {}                & {}              & $15.1\pm 0.1$    &  {}              & {}\\
AO~$0235+164$ & {}            & {}                & {}                  & $17.0\pm 0.1$    & {}              & {}\\
PKS~$0521-365$ & {}                & {}                & {}              & {}                 & {}              & {}\\
S5~$0716+714$ & $13.4\pm 0.1$   & {}                & $12.9\pm 0.1$ & $12.8\pm 0.1$    & $12.8\pm 0.1$   & {}\\
S5~$0836+710$ & $16.6\pm 0.1$   & {}                & $15.9\pm 0.1$ & $16.1\pm 0.1$    & {}              & $16.5\pm 0.1$ \\
Mkn~$421$ & {}                & {}                & {}              & {}                 & {}              & {}\\
{}           & {}                & {}                & {}              & {}                 & {}              & {}\\
{}           & {}                & {}                & {}              & {}                 & {}              & {}\\
{}           & {}                & {}                & {}              & {}                 & {}              & {}\\
{}           & {}                & {}                & {}              & {}                 & {}              & {}\\
{}           & {}                & {}                & {}              & {}                 & {}              & {}\\
PKS~$1127-145$ & {}            & {}                & {}              & $15.7\pm 0.1$    & $15.7\pm 0.1$ & $15.9\pm 0.1$ \\
ON~$231$ & $14.6\pm 0.1$   & $15.0\pm 0.1$   & $14.2\pm 0.1$ & {}                 & {}              & {} \\
3C~$273$ & $12.6\pm 0.1$   & $12.9\pm 0.1$   & $11.8\pm 0.1$ & $11.5\pm 0.1$    & $11.3\pm 0.1$ & $11.3\pm 0.1$ \\
{}           & $12.7\pm 0.1$   & $12.9\pm 0.1$   & $11.7\pm 0.1$ & {}                 & {}              & {} \\
{}           & {}                & {}                & {}              & $11.5\pm 0.1$    & $11.3\pm 0.1$ & $11.3\pm 0.1$\\
{}           & $12.7\pm 0.1$   & $12.9\pm 0.1$   & $11.8\pm 0.1$ & $11.5\pm 0.1$    & $11.3\pm 0.1$ & $11.3\pm 0.1$ \\
{}           & {}                & {}                & {}              & $11.3\pm 0.1$    & $11.1\pm 0.1$ & $11.1\pm 0.1$ \\
{}           & {}                & {}                & {}              & {}                 & {}              & {} \\
{}           & {}                & {}                & {}              & {}                 & {}              & {} \\
{}           & {}                & {}                & {}              & {}                 & {}              & {} \\
{}           & {}                & {}                & {}              & {}                 & {}              & {} \\
{}           & {}                & {}                & {}              & $11.4\pm 0.1$    & $11.3\pm 0.1$ & {} \\
{}           & {}                & {}                & {}              & {}                 & {}              & {} \\
{}           & {}                & {}                & {}              & {}                 & {}              & {} \\
{}           & {}                & {}                & {}              & {}                 & {}              & {} \\
{}           & {}                & {}                & {}              & {}                 & {}              & {} \\
{}           & $12.7\pm 0.1$   & $13.0\pm 0.1$   & $11.8\pm 0.1$ & $11.6\pm 0.1$    & $11.5\pm 0.1$ & $11.4\pm 0.1$ \\
Cen~A$^{*}$ & {}                & {}                & {}              & {}                 & {}              & {}  \\
PKS~$1334-127$ & {}            & {}                & {}              & $16.3\pm 0.1$    & $16.2\pm 0.1$  & {} \\
PKS~$1406-076$ & {}                & $19.5\pm 0.2$     & {}              & $18.6\pm 0.1$    & {}              & $19.7\pm 0.1$ \\
{}           & {}                & {}                & {}              & $19.4\pm 0.4$      & {}              & {} \\
NGC~$6251$ & {}                & $15.6\pm 0.1$   & $16.4\pm 0.1$  & $16.6\pm 0.1$   & {}              & {}  \\
PKS~$1830-211^{*}$ & {}                & {}                & {}              & {}                 & {}              & {} \\
{}           & {}                & {}                & {}              & {}                 & {}              & {} \\
{}           & {}                & {}                & {}              & {}                 & {}              & {} \\
PKS~$2155-304$ & {}                & {}                & {}              & {}                 & {}              & $12.5\pm 0.1$ \\
{}           & {}                & {}                & {}              & {}                 & {}                & $12.1\pm 0.1$ \\
{}           & {}                & {}                & {}              & {}                 & {}                & $12.2\pm 0.1$ \\
{}           & $13.2\pm 0.1$     & $13.4\pm 0.1$     & $12.4\pm 0.1$ & {}                 & {}              & {}           \\
{}           & $13.9\pm 0.1$     & $14.2\pm 0.1$     & $13.2\pm 0.1$ & {}               & {}              & $12.8\pm 0.1$ \\
{}           & $13.8\pm 0.1$ & $14.2\pm 0.1$   &  {}              & $13.1\pm 0.1$  & $12.9\pm 0.1$ & $12.9\pm 0.1$ \\
{}           & {}                & {}                & $12.9\pm 0.1$ & $12.7\pm 0.1$  & $12.6\pm 0.1$ & {} \\
{}           & $13.5\pm 0.1$ & {}                & $13.2\pm 0.1$ & {}                 & {}              & {}           \\
{}           & {}                & $14.3\pm 0.1$ & {}              & $12.7\pm 0.1$  & $12.6\pm 0.1$ & $12.6\pm 0.1$  \\
\hline
\end{tabular}
\begin{list}{}{}
\item[$^{*}$] Source beyond the capabilities of OM: too faint or too bright.
\end{list}
\label{tab:omdata}
\end{table*}

\emph{3EG~J0237+1635 (AO~0235+164):} 
The data set analyzed here refers to an observation performed in $2002$ and shows no signs of high background.
The \texttt{epatplot} task of \texttt{XMM SAS} shows a slight excess of double pixels events and a corresponding deficit 
of single pixel events (pile-up), that can be easily suppressed by removing the inner region with $5''$ radius. 
The fit can improve significantly by reducing the energy band to $1-10$~keV, that is by removing the energy band that
can be affected by the absorption and the fit reported in Tables~\ref{tab:srcdata} and \ref{tab:srcdata2} refer
to this case. The best fit is still obtained with the broken power law model with the absorption column fixed to the
Galactic value, although the $\Gamma_1$ is slightly harder than the above mentioned case.

\vskip 12pt

\emph{3EG~J0530-3626 (PKS~0521-365):} The blazar PKS~$0521-365$ has been associated with the EGRET source in the Second Catalog (Thompson et al. 1995), but
a stronger detection in the Cycle 4 placed this blazar outside the $99.99$\% probability contours. 
Sowards-Emmerd et al. (2004) suggested that the counterpart of 3EG~J$0530-3626$ source could be another radio source 
(PMN~J$0529-3555$) of unknown nature. This can be another case of a possible double source not resolved by EGRET 
(like, e.g. 3EG~J$0222+4253$), an interesting target worth observing with \emph{GLAST}.
In the present work, we keep as valid the association with PKS~$0521-365$. 
The \emph{XMM-Newton} observation is analyzed here for the first time. There is no evidence of high background,
but the \texttt{epatplot} task shows that the data are affected by pile-up. 
This blazar has a small jet ($6''$), visible in radio (see Tingay \& Edwards 2002 for a description of the parsec 
scale structure), and optical wavelenghts (Danziger et al. 1979). \emph{Chandra} detected a jet-like feature of $2''$-size,
spatially coincident with the optical and radio structure (Birkinshaw et al. 2002). Hardcastle et al.~(1999) and
Birkinshaw et al.~(2002) reported also the presence of extended emission that can be fitted with thermal plasma model
with $kT=1.6$~keV and $Z_{\odot}=0.05$. 
Given the size of the PSF of EPIC camera on board \emph{XMM-Newton} ($12''-15''$ HEW, see 
Jansen et al. 2001), the above~mentioned structures are not resolved. Attempts to fit the low
energy part with a thermal plasma model (\texttt{mekal} or \texttt{raymond} models in \texttt{xspec}, not reported in the 
Tables), resulted in a ${\tilde{\chi}^2}\approx 1.08$ improved with respect to the simple power law ($\Delta \chi^2=84$
for a decrease of $2$~dof), but still worse than the broken power law model. 
The broken power law model absorbed by the Galactic column provides the best fit to the present data, consistent
with the results obtained by \emph{Einstein}, \emph{EXOSAT} (Pian et al. 1996), \emph{ROSAT} (Pian et al. 1996, 
Hardcastle et al. 1999), \emph{BeppoSAX} (Tavecchio et al. 2002).

\vskip 12pt

\emph{3EG~J0721+7120 (S5~0716+714):} This BL Lac object is known for its extreme variability, also at intraday time scales 
(cf Wagner \& Witzel 1995),
and it is extensively monitored by optical and radio ground telescopes (see, e.g. Raiteri et al. 2003).
The present observation was performed simultaneously with the end of a ToO of \emph{INTEGRAL}, triggered after an
optical giant flare (Pian et al. 2005). The optical lightcurve started to increase at the end of March 2004, and
the trigger was activated on $27$ March. The \emph{INTEGRAL} ToO was performed from $2$ to $7$ April, when the 
source activity was already declining. The \emph{XMM-Newton} observation covered instead the period $4-5$~April. 
The X-ray lightcurve shows a clear decrease of the flux as the observation proceeded, confirming that the source was 
observed during the tail of the flare. 
More details on this \emph{XMM-Newton} data set are presented in a separate paper (Foschini et al., submitted). 

\vskip 12pt

\emph{3EG~J0845+7049 (S5~0836+710):} This flat-spectrum radio quasar has one of the highest redshifts among the objects in 
the present sample. The \emph{XMM-Newton} data, published here for the first time, show high background. 
Past analyses of \emph{ROSAT} and \emph{ASCA} data by Cappi et al. (1997) and \emph{BeppoSAX} data by Tavecchio et al. (2000) 
have shown a spectrum with a hard photon index ($\Gamma\approx 1.3$), with acceptable fit also with a broken power law,
but some inconsistencies in the value of the absorption. The present data are best fitted with a simple power law model
absorbed by the Galactic column plus an additional absorber at the redshift of the quasar with a value 
$(1.4\pm 0.3)\times 10^{21}$~cm$^{-2}$, consistent with the \emph{ASCA} and \emph{BeppoSAX} values. 
This is also consistent with \emph{Chandra} observations by Fang et al. (2001), who found $\Gamma=1.388\pm 0.012$ 
($0.5-8$~keV), but a slightly lower absorption $N_{\rm H}=(7.0\pm 1.2)\times 10^{20}$~cm$^{-2}$ (not redshifted). 
The photon index is also consistent with the value of $\Gamma=1.3\pm 0.3$ in the 
$20-100$~keV energy band measured by \emph{INTEGRAL} (Pian et al. 2005) and the value of $\Gamma = 1.1\pm 0.3$ measured
by \emph{CGRO} with OSSE and BATSE instruments (Malizia et al. 2000).
The broken power law gives also a good fit, although with no improvement with respect to the simple power law model, 
but does not require the additional absorber. 

\vskip 12pt
\emph{3EG~J1104+3809 (Mkn~421):} This is one of three sources in the present sample that has been extensively observed, being a calibration target
(the others are 3C~$273$ and PKS~$2155-304$). $32$ observations are present in the \emph{XMM-Newton} data archive, but
for sake of homogeneity in the present analysis, we selected only the $6$ observations with EPIC PN in small window mode.
We refer the reader to the several papers published with more detailed analysis of Mkn~$421$ data with different observing
modes (e.g. Brinkmann et al. 2001, 2003, Sembay et al. 2002, Ravasio et al. 2004).

\vskip 12pt
\emph{3EG~J1134-1530 (PKS~1127-145):} This source is in the list of Gigahertz-Peaked Sources (GPS) by 
Stanghellini et al. (1998), Stanghellini (2003). The present \emph{XMM-Newton} data set has never been published. 
It shows high background, but, once cleaned of soft-proton flares, it is possible to extract useful information. 
This source also has a $30''$-sized jet, observed in X-rays with \emph{Chandra} 
(Siemiginowska et al. 2002), but that is still too small for the PSF size of EPIC. 

\vskip 12pt
\emph{3EG~J1222+2841 (ON~231):} This is another EGRET source that could be composed of two or more contributions: 
indeed, although ON~$231$ is outside the global $99$\% probability contours of 3EG~J$1222+2841$, there is a strong 
association with the emission at $E>1$~GeV (Lamb \& Macomb 1997). 
The present \emph{XMM-Newton} data set has never been analysed and shows evidence of high background towards the end
of the observation. The broken power law model provides the best fit to the data. An absorption in addition to the
Galactic column is marginally detected. The values are consistent with, although slightly different to
the fit to the \emph{BeppoSAX} data of an observation in $1998$, when ON~$231$ was in outburst 
(Tagliaferri et al. 2000).

\vskip 12pt
\emph{3EG~J1229+0210 (3C~273):} For historical reasons, 3C~$273$ is one of the most observed sources in the sky, 
and \emph{XMM-Newton} spent a lot of time observing this quasar (3C~$273$ is also a calibration source). 
We refer the reader to the works by Molendi \& Sembay (2003), Courvoisier et al. (2003) and Page et al. (2004) for 
more details in the analysis of the \emph{XMM-Newton} data sets. There is a general agreement between the analysis 
presented here and the above cited works: just to mention one case (ObsID $0136550501$), for the fit with a simple 
power law model with the Galactic absorption column in the $3-10$~keV energy band, Courvoisier et al. (2003) 
reported $\Gamma=1.74\pm 0.03$, Page et al. (2004) found $\Gamma=1.78\pm 0.02$, and the value in the present work 
is $\Gamma=1.79\pm 0.06$.
However, small differences in the procedures should be noted. Indeed, a small excess of double pixel events and a 
corresponding deficit of single pixel events can be observed with \texttt{epatplot} task at high energies, 
specifically above $8$~keV, as already indicated by Molendi \& Sembay (2003). The problem is resolved by removing 
the inner region of $8''$ radius: the fit improves with negligible changes in the spectral parameters. For example, 
let us consider the data of the ObsID $0112770101$ in the $3-10$~keV energy band fit with a simple power law model 
with Galactic absorption. By using a circular region with $40''$ radius, the best fit gives these values: 
$\Gamma=1.69\pm 0.05$ and normalization
$0.033\pm 0.003$~ph~cm$^{-2}$~s$^{-1}$~keV$^{-1}$ for ${\tilde{\chi}^2}=1.08$ and $588$ dof. This is to be compared with
$\Gamma=1.73\pm 0.07$ and normalization $0.036\pm 0.004$~ph~cm$^{-2}$~s$^{-1}$~keV$^{-1}$ for ${\tilde{\chi}^2}=0.97$ and 
$315$ dof in the case of annular region of extraction with inner radius $8''$ and external radius $40''$. Therefore,
since the fit improves significantly with a small increase of the error bars, the annular region of extraction has been
used in the data set analyzed here. 
An alternative model is a power law plus a blackbody (Table~\ref{tab:srcdata3}), where the thermal component could have 
a physical origin and be the hard tail of a Comptonized accretion disk. However, the values found in the present work 
are well above the value of $54_{-4}^{+6}$~eV found with \emph{BeppoSAX}, but with a more complex model 
(Grandi \& Palumbo 2004). On the other hand, 3C~$273$ appeared to be in a different state
when observed with \emph{BeppoSAX} and with \emph{XMM-Newton} (see also Page et al. 2004).

\vskip 12pt
\emph{3EG~J1324-4314 (Cen~A):} Centaurus~A is the nearest AGN in the sky ($d=3.84$~Mpc) and has been observed twice 
by \emph{XMM-Newton} with a delay of one year between the two observations. An analysis of these data has been 
published by Evans et al. (2004) and we refer the reader to that paper for more details, particularly for the extranuclear
environment. But, since the purpose of the present work is to study the continuum, we performed the fit in the 
$4-10$~keV energy band, to avoid the complex features in the low energy part of the spectrum. Moreover, to guarantee 
a good approximation of the continuum, we ignored the energy band $6-8$~keV, which is affected by prominent emission lines. 
After having fixed the power law, the energy band of the iron complex is restored and one or two Gaussian emission 
lines are added to the model to complete the fit (Table~\ref{tab:srcdata4}). Both observations require a Gaussian 
emission line from FeK$\alpha$, but a wing toward the high energy is present. The addition of another large line with 
centroid at $6.8$~keV determines an improvement in the fit of the ObsID~$0093650201$ (although the centroid is not well 
constrained), but not in the ObsID~$0093650301$ (it should be noted that this ObsID has less statistical power, because the 
MOS2 data are not useful). The iron complex found here is partially in agreement with 
the \emph{BeppoSAX} results obtained by Grandi et al. (2003): the discrepancies refer to the FeK$\beta$ line at 
$7.1$~keV, that is not required by the present data sets. The variability of the neutral iron line found by 
Grandi et al. (2003) is confirmed also by the present data: the line flux changed significantly between the two 
observations (spaced by 1 year), but increasing with the source flux increases, the opposite of what has been found 
in \emph{BeppoSAX} data. However, two points temporally spaced by one year are not sufficient to claim 
different behaviour. 

\vskip 12pt
\emph{3EG~J1339-1419 (PKS~1334-127):} This flat-spectrum radio quasar has been poorly observed in hard X-rays: 
the only available observations are with \emph{ROSAT} and \emph{Einstein} (see, e.g., Maraschi et al. 1995a). 
The present analysis is also the first look at the hard X-ray emission ($E>4$~keV) of this source. 
A simple power law model, with absorption in excess of the Galactic column, provides the best fit. The photon index
is in the middle of the values from \emph{Einstein} and \emph{ROSAT} (Maraschi et al. 1995a), and consistent with
both within the $90$\% confidence level.

\vskip 12pt
\emph{3EG~J1409-0745 (PKS~1406-076):} This source was never studied in X-rays: \emph{ROSAT} observation resulted 
only in an upper limit ($2\sigma$) with $F_{0.1-2.4\: \rm keV}<2.35\times 10^{-13}$~erg~cm$^{-2}$~s$^{-1}$ 
(Siebert et al. 1998). \emph{XMM-Newton} observed this blazar twice, with the second observation about one 
month after the first one, and the fluxes in the \emph{ROSAT} energy band were $4.2$ and 
$3.9\times 10^{-13}$~erg~cm$^{-2}$~s$^{-1}$, respectively. 
In both observations, the source was best fitted with a simple power law with $\Gamma \approx 1.6$ and no 
additional absorption.

\vskip 12pt
\emph{3EG~J1621+8203 (NGC~6251):} This EGRET source has been associated by Mukherjee et al.~(2002) 
with the nearby FRI radio galaxy NGC~$6251$ ($z=0.02471$). Later studies 
supported this conclusion, e.g. Sowards-Emmerd et al. (2003), Chiaberge et al. (2003), Guainazzi et al. (2003), 
Foschini et al. (2005). There is one \emph{XMM-Newton} pointing available which lasted $50$~ks, but heavy 
contamination with soft-proton flares strongly reduced the effective exposure on the three detectors (see 
Table~\ref{tab:log}). The best fit model is an absorbed broken power law model. The addition of a thermal plasma 
to the single power law does not provide an improvement with respect to the broken power law model. 
This data set has been analyzed by Gliozzi et al. (2004) and Sambruna et al. (2004), 
but there are some discrepancies with the present analysis (mainly the presence of the FeK$\alpha$ emission line) 
likely to be attributed to a different cleaning of soft-proton flares. In the present analysis, the addition of
a narrow ($0.1$~keV) or broad ($0.5$~keV) line at $6.4$~keV, determines a worsening of the fit both with respect
to the single and the broken power law model. On the other hand, the present results are consistent with \emph{ASCA} 
and \emph{BeppoSAX} observations analyzed by Chiaberge et al. (2003) and Guainazzi et al. (2003) that found no 
indication of any emission line of the iron complex. An improvement with respect to the single power law is obtained by
adding the \texttt{mekal} model, but the fit remains always worse than the broken power law model.
The parameters of the single power law plus thermal plasma model are: $N_{\rm H}=(1.16\pm 0.08)\times 10^{21}$~cm$^{-2}$, 
$kT=0.58_{-0.13}^{+0.16}$~keV, $\Gamma=1.91\pm 0.03$ for ${\tilde{\chi}^2}=1.02$ and $810$ dof. The observed flux in
the $0.4-10$~keV energy band is $5.81\times 10^{-12}$~erg~cm$^{-2}$~s$^{-1}$.

\vskip 12pt
\emph{3EG~J$1830-2110$ (PKS~$1830-211$):} This is the highest redshift blazar in the present sample ($z=2.507$) and
is gravitationally lensed by an intervening galaxy at $z=0.886$ (Wiklind \& Combes 1996, Lidman et al. 1999, 
Courbin et al. 2002). It was observed in the past in X-ray by \emph{ROSAT} (Mathur \& Nair 1997), \emph{ASCA} 
(Oshima et al. 2001) and \emph{Chandra} (De Rosa et al. 2005). The present data set has not yet been published and
two observations of the three available are affected by high background.

\vskip 12pt
\emph{3EG~J2158-3023 (PKS~2155-304):} This BL Lac object is the third calibration source in the present sample and, for this reason, it has been observed
several times. Most of the present data set have been already published (Edelson et al. 2001, Maraschi et al. 2002, 
Cagnoni et al., 2004, Zhang et al. 2005), where it is possible to find more detailed analyses, particularly with reference 
to timing properties. This source, being one of the brightest in the X-ray sky, is also used in the search for the
warm-hot intergalactic medium (WHIM) and a local absorber has been detected at $21.59\, \AA$ ($\approx 0.57$~keV) with
the Reflection Grating Spectrometers (RGS) on board \emph{XMM-Newton} (Cagnoni et al. 2004). However, since the study of 
this type of feature is outside the purpose of the present work, the spectrum of PKS~$2155-304$ has been fitted in the 
$0.6-10$~keV energy range and then the flux has been extrapolated to $0.4$~keV. 

\clearpage

\end{document}